\documentclass[a4paper,fleqn,usenatbib]{mnras}

\usepackage[T1]{fontenc}
\usepackage{ae,aecompl}

\usepackage{longtable}
\usepackage{graphicx}	
\usepackage{amsmath}	
\usepackage{amssymb}	
\usepackage[T1]{fontenc}
\usepackage[flushleft]{threeparttable}
\usepackage{lscape}
\usepackage{multicol}
\usepackage{xcolor}
\usepackage{upgreek}
\usepackage{breqn}
\usepackage[flushleft]{threeparttable}

\usepackage{ragged2e}
\newcommand{\microns}{\,$\upmu\mathrm{m}$}

\definecolor{GeorgeBlue}{RGB}{0,128,128}

\definecolor{referee}{RGB}{255,0,0}

\title[BEARS I: Redshifts of bright lensed H-ATLAS galaxies]{The Bright Extragalactic ALMA Redshift Survey (BEARS) I: redshifts of bright gravitationally-lensed galaxies from the \textit{Herschel} ATLAS}

\author[S.A.Urquhart et al.]{S. A. Urquhart$^{1}$\thanks{E-mail: sheona.urquhart@open.ac.uk},
G. J. Bendo$^{2}$,
S. Serjeant$^{1}$,
T. Bakx$^{3,36}$, 
M. Hagimoto$^{36}$,
P. Cox$^{4}$,
R. Neri$^{5}$,
\newauthor
M. Lehnert$^{4,26}$, 
C. Sedgwick$^{1}$,
C. Weiner$^{1}$,
H. Dannerbauer$^{9, 10}$, 
A. Amvrosiadis$^{11}$,
P. Andreani$^{30}$,
\newauthor
A.J. Baker$^{6}$, 
A. Beelen$^{7}$, 
S. Berta$^{1}$,
E. Borsato$^{41}$,
V. Buat$^{7}$, 
K.M. Butler $^{24}$,
A. Cooray$^{8}$, 
G. De Zotti$^{28}$,
\newauthor
L. Dunne$^{11}$, 
S. Dye$^{12}$, 
S. Eales$^{11}$, 
A. Enia$^{28}$,
L. Fan$^{34}$,
R. Gavazzi$^{4,27}$, 
J. Gonz\'alez-Nuevo$^{31, 40}$
\newauthor
A.I. Harris$^{13}$, 
C.N. Herrera$^{5}$, 
D. Hughes$^{14}$, 
D. Ismail$^{7}$,
R. Ivison$^{15}$, 
S. Jin$^{9, 10}$, 
B. Jones$^{39}$,
K. Kohno$^{37, 38}$,
\newauthor
M. Krips$^{5}$, 
G.Lagache$^{7}$, 
L. Marchetti$^{17, 33}$, 
M. Massardi$^{16,33}$,
H. Messias$^{19}$, 
M. Negrello$^{11}$, 
A. Omont$^{4}$, 
\newauthor
I. Perez-Fournon,$^{9}$ 
D.A. Riechers$^{21}$, 
D. Scott$^{32}$,
M.W.L. Smith$^{11}$,
F. Stanley$^{4}$,
Y. Tamura$^{36}$,
P. Temi$^{29}$,
\newauthor
C. Vlahakis$^{22}$,
A.Wei{\ss}$^{23}$, 
P. van der Werf$^{24}$, 
A. Verma$^{35}$,
C. Yang$^{25}$ 
and A.J. Young$^{6}$ \\
\\
$^{1}$School of Physical Sciences, The Open University, Milton Keynes, MK7 6AA, UK\\
$^{2}$UK ALMA Regional Centre Node, Jodrell Bank Centre for Astrophysics,
    Department of Physics and Astronomy, \\ The University of Manchester, 
    Oxford Road, Manchester M13 9PL, United Kingdom\\
$^{3}$National Astronomical Observatory of Japan, 2-21-1, Osawa, Mitaka, Tokyo 181-8588, Japan\\
$^{4}$Sorbonne Universit\'e, UPMC Universit\'e Paris 6 and CNRS, UMR 7095, Institut d'Astrophysique de Paris,\\ 
98bis boulevard Arago, 75014 Paris, France \\
$^{5}$Institut de Radioastronomie Millim\'etrique (IRAM), 300 rue de la Piscine, 38400 Saint-Martin-d'H\'eres, France\\
$^{6}$ Department of Physics and Astronomy, \\Rutgers, The State University of New Jersey, 136 Frelinghuysen Road, Piscataway, NJ, 08854-8019, USA \\
$^{7}$Aix-Marseille Universit\`e, CNRS and CNES, Laboratoire d'Astrophysique de Marseille, 38, rue Fr\`ed\`eric Joliot-Curie,\\ 
13388 Marseille, France\\
$^{8}$University of California Irvine, Physics and Astronomy, FRH 2174, Irvine CA 92697, USA\\
$^{9}$Instituto Astrof\'isica de Canarias (IAC), E-38205 La Laguna, Tenerife, Spain \\
$^{10}$Universidad de la Laguna, Dpto. Astrof\'isica, E-38206 La Laguna, Tenerife, Spain\\
$^{11}$School of Physics and Astronomy, Cardiff University, The Parade, Cardiff, CF24 3AA, UK\\
$^{12}$School of Physics and Astronomy, University of Nottingham, University Park, Nottingham, NG7 2RD, UK\\
$^{13}$Department of Astronomy, University of Maryland, College Park, MD 20742, USA\\
$^{14}$Instituto Nacional de Astrof\`isica, \`Optica y Electr\`onica, Astrophysics Department, Apdo 51 y 216, Tonantzintla,\\ 
Puebla 72000, Mexico\\
$^{15}$European Southern Observatory, Karl-Schwarzschild-Strasse 2, D-85748 Garching, Germany\\
$^{16}$SISSA, Via Bonomea 265, 34136 Trieste, Italy\\
$^{17}$Department of Astronomy, University of Cape Town, 7701 Rondebosch, Cape Town, South Africa\\
$^{18}$Department of Physics and Astronomy, University of the Western Cape, Private Bag X17, Bellville 7535, Cape Town, South Africa \\
$^{19}$Instituo de Astrof\`isica e Ci\^encias do Espa\c co, Tapada da Ajuda, Edif\`icio Leste, 1349-018 Lisboa, Portugal\\
$^{20}$Department of Astronomy, Cornell University, Space Sciences Building, Ithaca, New York (NY) 14853, USA \\
$^{21}$I. Physikalisches Institut, Universit\"at zu K\"oln, Z\"ulpicher
Strasse 77, D-50937 K\"oln, Germany
\\
$^{22}$National Radio Astronomy Observatory, 520 Edgemont Road, Charlottesville VA 22903, USA\\
$^{23}$Max-Planck-Institut f\"ur Radioastronomie, Auf dem H\"ugel 69, 53121 Bonn, Germany\\
$^{24}$Leiden University, Leiden Observatory, PO Box 9513, 2300 RA Leiden, The Netherlands\\
$^{25}$European Southern Observatory, Alonso de Co\`ordova 3107, Casilla 19001, Vitacura, Santiago, Chile\\
$^{26}$ Universit\'{e} Lyon 1, ENS de Lyon, CNRS UMR5574, Centre de Recherche Astrophysique de Lyon, F-69230 Saint-Genis-Laval, France\\
$^{27}$ Institute of Astronomy, University of Cambridge, Madingley Road, Cambridge CB3 0HA, UK \\
$^{28}$ INAF-Osservatorio Astronomico di Padova, Vicolo dell'Osservatorio 5, I-35122 Padova, Italy \\
$^{29}$ Astrophysics Branch, NASA Ames Research Center, Moffett Field, CA, USA \\
$^{30}$ 7 AIM, CEA, CNRS, Universit\'e Paris-Saclay, Universit\'e Paris Diderot, Sorbonne Paris Cit\'e, Observatoire de Paris, \\
PSL University, 91191 Gif-sur-Yvette Cedex, France \\
$^{31}$ Departamento de Fisica, Universidad de Oviedo, C. Federico Garica Lorca 18, 33007, Oviedo, Spain\\
$^{32}$ Department of Physics and Astronomy, University of British Columbia, 6224 Agricultural Road, Vancouver, BC, V6T 1Z1 \\
$^{33}$ INAF, Instituto di Radioastronomia-Italian ARC, Via Piero Gobetti 101, 40129, Bologna, Italy \\
$^{34}$ Shandong Key Laboratory of Optical Astronomy and Solar-Terrestrial Environment, Institute of Space Sciences, Shandong University, \\
Weihai, Shandong, 264209, China \\
$^{35}$ Astrophysics, Department of Physics, Keble Road, Oxford, OX1 3RH, UK \\
$^{36}$ Division of Particle and Astrophysical Science, Graduate School of Science, Nagoya University, Nagoya 464-8602, Japan\\
$^{37}$ Institute of Astronomy, Graduate School of Science, The University of Tokyo, 2-21-1 Osawa, Mitaka, Tokyo 181-0015, Japan  \\
$^{38}$ Research Center for Early Universe, Graduate School of Science, University of Tokyo, 7-3-1 Hongo, Bunkyo-ku, Tokyo 113-0033, Japan\\
$^{39}$ Jodrell Bank Centre for Astrophysics, School of Natural Sciences, The University of Manchester, Manchester, M13 9PL, UK \\
$^{40}$ Instituto Universitario de Ciencias y Tecnologias Espaciales de Asturias (ICTEA), C. Independencia 13, 33004, Oviedo, Spain \\
$^{41}$ Dipartimento di Fisica e Astronomia "G. Galilei", Universit\`a di Padova, vicolo dell'Osservatorio 3, I-35122 Padova, Italy\\ 
}
\date{Accepted XXX. Received YYY; in original form ZZZ}

\pubyear{2021}

\begin{document}
\label{firstpage}
\pagerange{\pageref{firstpage}--\pageref{lastpage}}
\maketitle

\begin{abstract}
We present spectroscopic measurements for 71 galaxies associated with 62 of the brightest high-redshift submillimeter sources from the Southern fields of the {\it Herschel} Astrophysical Terahertz Large Area Survey (H-ATLAS), while targeting 85 sources which resolved into 142.  We have obtained robust redshift measurements for all sources using the 12-m Array and an efficient tuning of ALMA to optimise its use as a redshift hunter, with 73 per cent of the sources having a robust redshift identification.  Nine of these redshift identifications also rely on observations from the Atacama Compact Array. The spectroscopic redshifts span a range $1.41<z<4.53$ with a mean value of 2.75, and the CO emission line full-width at half-maxima range between $\rm 110\,km\,s^{-1} < FWHM < 1290\,km\,s^{-1}$ with a mean value of $\sim$500\,km\,s$^{-1}$, in line with other high-$z$ samples.  The derived CO(1-0) luminosity is significantly elevated relative to line-width to CO(1-0) luminosity scaling relation, which is suggestive of lensing magnification across our sources. In fact, the distribution of magnification factors inferred from the CO equivalent widths is consistent with expectations from galaxy-galaxy lensing models, though there is a hint of an excess at large magnifications that may be attributable to the additional lensing optical depth from galaxy groups or clusters. 
\end{abstract}




\section{Introduction}\label{sec:introduction}

Dusty sub-millimetre galaxies (SMGs) were particularly important contributors to the overall star formation budget in the early Universe \citep[e.g.,][]{hodge2020}.  With total infrared luminosities exceeding 10$^{12}$\,L$_{\odot}$, SMGs reach the limit of `maximum starburst' with star formation rates of 1000\,M$_{\odot}$yr$^{-1}$ or more {\citep[e.g.,][]{RowanRobinson2016}}.  While their exact nature is still debated \citep[e.g.,][]{narayanan2015}, many of them are likely to be mergers \citep[e.g.,][]{engel2010,tacconi2008}, although the general population is likely more diverse \citep[e.g.,][]{lapi2011}.  Compared to local ultra-luminous infrared galaxies (ULIRGs), SMGs at the peak of cosmic evolution ($z = 1.5-4$) are orders of magnitude more numerous and luminous.  Having a median redshift of $z \sim 2.5$  \citep[e.g.,][]{danielson2017}, the SMG population significantly contributes to the peak of the cosmic star-formation rate density at $z \sim 2-3$ \citep{madau2014} and therefore plays a critical role in the history of cosmic star formation and the physical processes driving the most extreme phases of galaxy formation and evolution \citep[e.g.,][]{swinbank2010,walter2012}.

Large area submm/mm-wave surveys have proven transformative for extragalactic astronomy, such as the Atacama Cosmology Telescope \citep[e.g.,][]{marsden14}, Planck \citep[e.g.,][]{Harrington2021}, and the South Pole Telescope \citep[e.g.,][]{reuter20}. In particular, the {\it Herschel Space Observatory} has increased the number of known SMGs from hundreds to hundreds of thousands through a series of surveys, specifically: the {\it Herschel} Astrophysical Terahertz Large Area Survey \citep[H-ATLAS;][]{eales2010} and the {\it Herschel} Multi-tiered Extragalactic Survey  \citep[HerMES;][]{oliver2012}, with a total area of over 1000\,deg$^{2}$.  The surface density of unlensed sources drops quickly at the 500$\mu$m flux density S$_{500\mu m}\gtrsim 100$\,mJy, and objects above this threshold are almost all gravitationally-magnified by a foreground galaxy or galaxy cluster.  These large-area surveys have therefore enabled the detection of numerous SMGs that are amongst the brightest in the sky, containing a large fraction of the rare high-redshift strongly lensed SMGs \citep{negrello2010,wardlow2013,nayyeri2016,bakx18} and hyper-luminous infrared galaxies (HyLIRGS; $L_{\rm FIR}\geq 10^{13}$\,L$_{\odot}$, e.g., \citealp{ivison2013}; \citealp{fu2013}).  Strong gravitational lensing allows access to populations that would otherwise be inaccessibly faint, and the angular magnification permits detailed $\sim$100\,pc resolution analysis of star formation in follow-up observations.  Indeed, the background submm-bright galaxies are ideal targets for sub/millimeter-wave interferometers.

Precise redshift measurements are essential for determining many fundamental properties of SMGs, and for measuring their clustering power spectrum.  Photometric redshifts are only approximate (due to the degeneracy with dust temperature, e.g., \citealp{blain1999}), therefore 
one has to rely on spectroscopic methods \citep{casey2012} that are more expensive in terms of telescope time. Optical/near-infrared ground-based spectroscopic redshift campaigns on 8m-class facilities only succeed for a minority of sources for which precise positions are available through their faint radio emission \citep[e.g.,][]{ivison1998}, when known, but for most bright SMGs the very high dust extinction prevents optical/near-infrared redshift determination, particularly at the highest redshifts \citep[e.g.,][]{chapman2015}. However, redshifted carbon monoxide (CO) emission lines are observable with sub-millimetre and millimetre-wave spectroscopy. These emission lines are unobscured by dust exinction and are directly attributable to the sub/mm sources.  

The increased bandwidths of the receivers operating at sub/mm have made sub/mm spectroscopy technically feasible for SMG redshift determinations, despite the SMG population having been detected in the continuum for decades.  Early successes include the Cosmic Eyelash SMMJ14009+0252 \citep{weiss2009,swinbank2010} at the 30-meter telescope, and HDF850.1 at the Plateau de Bure interferometer \citep{walter2012}.  The availability of various broadband instruments on the Green Bank Telescope \citep{harris2012}, CARMA \citep{riechers2011} and with the Caltech Sub-millimetre Observatory \citep{lupu2012} enabled the measurement of redshifts for very bright sources selected from the {\it Herschel} surveys.

More recently, using the Atacama Large Millimeter/submillimeter Array (ALMA), \cite{weiss2013} presented a redshift survey for 23 strongly lensed dusty star-forming galaxies selected from the South Pole Telescope (SPT) 2500 deg$^{2}$ survey.  This work was followed by further ALMA observations yielding reliable measurements for redshifts of an additional 15 high-redshift luminous galaxies from the SPT \citep{strandet2016,reuter20} and provided a larger set of redshifts from the SPT sample, totalling 81 galaxies with median redshift of $z$=3.9, selected with two flux limits, at 1.4\,mm and $870$\microns (see \cite{reuter20} for more details). The longer-wavelength selection increases the high redshift tail \cite{Marrone2018}. Similarly, \cite{neri2019} measured the redshifts of 13 bright galaxies detected in H-ATLAS with S$_{500}\geq 80$\,mJy, deriving robust spectroscopic redshifts for 12 individual sources, based on the detection of at least two emission lines, having a median redshift of $z$=2.9.  
Following this successful pilot study, a large comprehensive survey (z-GAL; PIs: P. Cox, T. Bakx and H. Dannerbauer) has recently been completed with NOEMA. Reliable redshifts were derived for all 126 bright Herschel-selected SMGs with 500 mu fluxes > 80 mJy that were selected from the H-ATLAS and HerMES fields in the Northern and equatorial planes. The results of this large programme will soon be reported in a series of dedicated papers.

Here, we present robust spectroscopic redshift measurements from the 12-m Array obtained in ALMA Cycle 7 and from the ACA (Atacama Compact Array) in ALMA Cycles 4 and 6 for 71 galaxies: Bright Extragalactic ALMA Redshift Survey (BEARS).  The results from this redshift campaign enable a wide range of follow-up observations, such as using emission lines to map the dynamics of dusty galaxies with the benefit of strong lensing angular magnifications, determining the physical properties of the sources' interstellar media (e.g., ionisation state, density), and conducting spectroscopic searches for companions. Paper II in this series (Bendo et al. in prep.) will present continuum measurements from these data as well as analyses of spectral energy distributions, while paper III (Hagimoto et al. in prep.) will present inferences from the CO ladder and composite spectrum. 

We structure the paper as follows.  In Section \ref{sec:data}, we describe the sample selection and ALMA observations carried out in Cycles 4 and 6 using the ACA and Cycle 7 using the 12-m Array.  In Section \ref{sec:results} we describe how we obtain our redshift measurements using single and multiple emission lines (where detected) and how these compare with literature.  Section \ref{sec:discussion} discusses our redshift distribution, the comparison to other surveys, the correlation between line luminosity and velocity width, and the potential for differential magnification to affect the interpretation of our results. Finally, Section \ref{sec:conclusions} presents our conclusions.

We adopt a spatially flat $\Lambda$CDM cosmology throughout this paper with $H_{0}=67.4$\,km\,s$^{-1}$\,Mpc$^{-1}$ and $\Omega_{\rm M}=0.315$ \citep{Planck2018parameters}.

\section{Data}\label{sec:data}
\subsection{Sample selection}

\begin{figure}
    \centering
    \includegraphics[width=\linewidth]{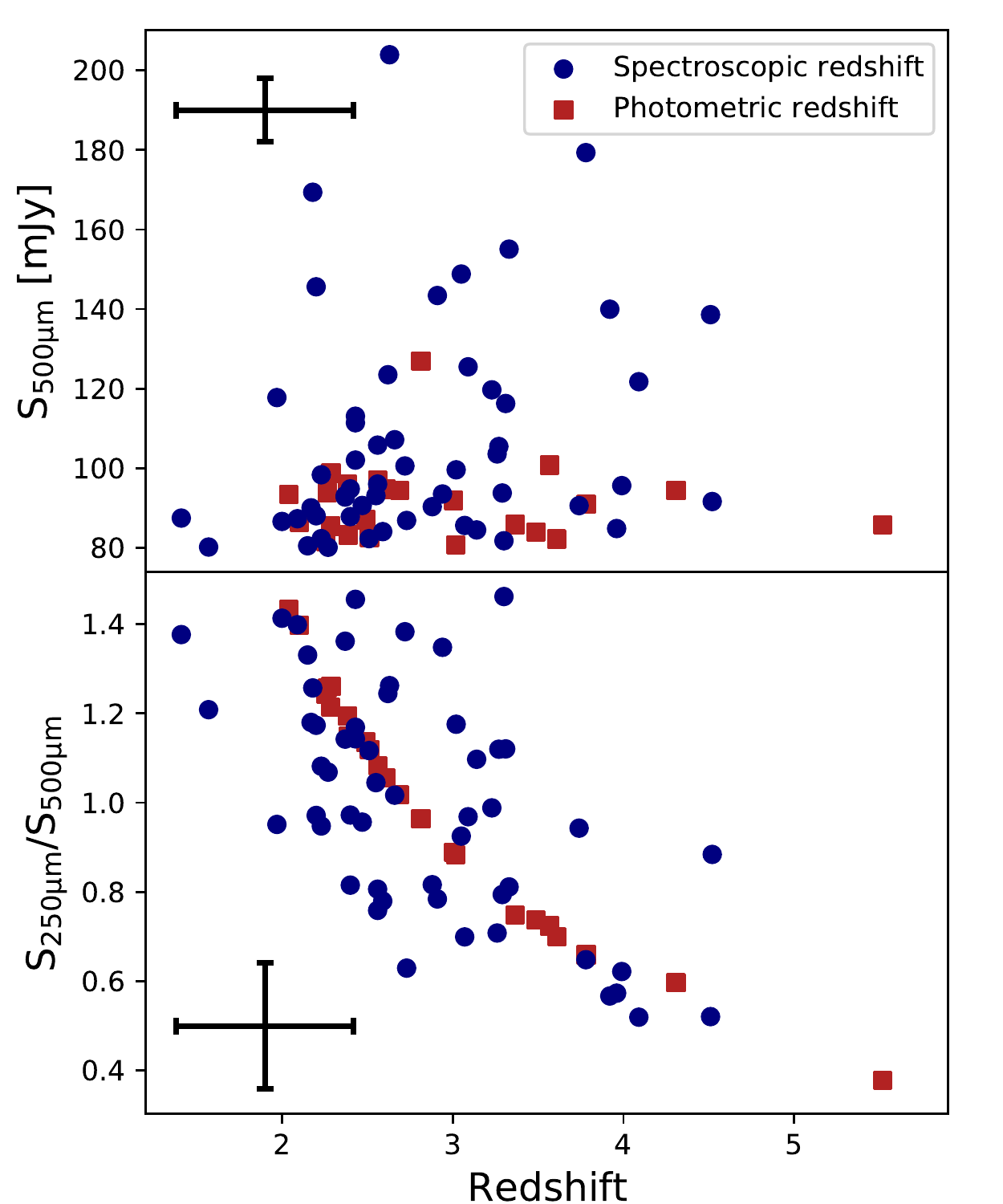}
    \caption{{\it Top panel:} 500\,$\mu$m flux density against redshift. Sources with spectroscopic redshifts are shown as {\it blue dots}, while sources with only photometric redshifts are shown as {\it red squares}. Sources are selected with 500\microns{} greater than 80\,mJy. We show the typical uncertainties on these values in the top-left of the figure, assuming a $z_{
    \rm phot} = 3$. For sources with multiple galaxies at different redshifts, we use the average spectroscopic redshift. {\it Bottom panel:} Submm colours and redshifts of our sample, illustrating the photometric selection function above $z_{\rm phot}=2$. Typical uncertainties in the colour and photometric redshift estimates are shown in the bottom-left of the graph.}
    \label{fig:colour_selection}
\end{figure}

Our targets are taken from the {\it Herschel} Astrophysical Terahertz Large Area Survey \citep[H-ATLAS,][]{eales2010}.  H-ATLAS was the largest open-time key project on {\it Herschel} in terms of time awarded and legacy catalogue size, covering 550 deg$^{2}$, which was by far the widest area of any extragalactic {\it Herschel} survey.  H-ATLAS mapped 150 deg$^{2}$  in the Northern Galactic Pole (NGP), three equatorial fields each of 36 deg$^{2}$ covering the Galaxy and Mass Assembly (GAMA) survey at RAs of 9h, 12h and 15h and two Southern Galactic Pole (SGP) fields of 102 deg$^{2}$ and 160 deg$^{2}$ respectively. 

Our sample selection for this paper is based on the criteria set out in \citet{bakx18}:
\begin{itemize}
    \item $500$\microns\ flux density $\geq80$\,mJy;
    \item Lack of cross-identification with known blazars or bright local galaxies (following the gravitational lens selection technique of \citealp{negrello2010}); 
    \item Photometric redshift estimate of $z_{\rm phot}\geq2$ based on the {\it Herschel} SPIRE flux densities at $250$\microns, 350\microns\ and 500\microns; 
    \item Location in the H-ATLAS South Galactic Pole field. 
\end{itemize}

The photometric redshift estimates were derived using the two-temperature modified black-body template from \cite{pearson13}. In total, they find 209 sources, of which 88 were located in the SGP field, three of which already had spectroscopic redshifts at the start of our redshift campaign.

The spectral line survey was originally started as a pathfinder experiment with the ACA in Band 3 in Cycle 4 (in programme 2016.2.00133.S), and these Band 3 ACA observations were continued in Cycle 6 (in programme 2018.1.00804.S).  
An analysis after the Cycle 6 observations determined that more reliable redshifts could be measured by including Band 4 observations.  This led to observations with the ALMA 12~m Array that covered all 85 fields in Band 4 as well as Band 3 (in programme 2019.1.01477.S) where emission lines were not already obvious.

All the sources with redshifts have band\,4 observations using the ALMA 12\,m array, with 74 sources also having band\,3 observations using the 12\,m array. Eleven sources, instead, rely on ACA observations in band 3. We find spectroscopic redshifts for one or multiple sources in 62 of the 85 {\it Herschel} fields. As reported by Bendo et al. in prep, we find 142 individual galaxies in the ALMA images, and here we report the 71 galaxies with spectroscopic redshifts.
The targets are as presented in Table \ref{tab:data_table} (see Table A1 for integrated flux densities).

Figure~\ref{fig:colour_selection} shows the submm colours of the 85 selected sources, and illustrates the selection function resulting from the flux limit and photometric redshift constraint.  Of the 11 sources that rely on ACA data in band 3, 9 resulted in robust spectroscopic redshifts. This is a similar success rate as with the full baseline, and as such, we do not expect significant differences between the sources using ALMA full baseline array and ACA-dependent data. Note that the 2\,mm and 3\,mm integrated continuum flux density limits in our ALMA observations do not affect our sample selection in any way, because all targets were detected with continuum signal to noise ratios in excess of 8 at 3\,mm and 10 at 2\,mm (see Bendo et al., in prep.). This is also evident from Figure~\ref{fig:colour_selection} in which no trends in redshift determination with colour or redshift are apparent.

\onecolumn
\setlength\LTcapwidth{\linewidth}
\begin{longtable}{cccccccc} 
\caption{Sources with robust spectroscopic redshifts.  (1) H-ATLAS source ID; (2) HerBS ID \citep{bakx18} where available; (3) number of continuum sources; (4) source designation (section 2.1); (5) and (6) RA and Dec coordinates respectively; (7) spectroscopic redshifts obtained in this work; and (8) VIKING derived lens redshift from \citet{bakx2020}.
} 
\label{tab:data_table}  \\
\hline
\multicolumn{1}{c}{H-ATLAS ID} 		 & \multicolumn{1}{c}{HerBS ID} 		 & \multicolumn{1}{c}{Number} 		 & \multicolumn{1}{c}{Source} 		  & \multicolumn{2}{c}{Coordinates (J2000)} 		 
& \multicolumn{1}{c}{{$z_{\rm spec}$}}  		 & \multicolumn{1}{c}{$z_{\rm lens}$} \\

\multicolumn{1}{c}{ } 		 & \multicolumn{1}{c}{ } 		 & \multicolumn{1}{c}{of sources} 		 & \multicolumn{1}{c}{designation} 		  & \multicolumn{1}{c}{RA} 	 & \multicolumn{1}{c}{Dec} 				 & \multicolumn{1}{c}{ }  & \multicolumn{1}{c}{ }  \\	
\hline \endhead	
\hline \multicolumn{8}{r}{\textit{Continued on next page}} \\
\endfoot
\hline
\endlastfoot
J012407.4$-$281434&11&1&&01:24:07.50&$-$28:14:34.7&2.631\\
J013840.5$-$281856&14&1&&01:38:40.41&$-$28:18:57.5&3.782\\
\footnotemark[1]J232419.8$-$323927&18&1&&23:24:19.82&$-$32:39:26.5&2.182&0.647\\
\footnotemark[1]J234418.1$-$303936&21&2&[A+B]&23:44:18.11&$-$30:39:38.9&3.323\\
J002624.8$-$341738&22&2&A&00:26:24.99&$-$34:17:38.1&3.050\\
J004736.0$-$272951&24&1&&00:47:36.09&$-$27:29:52.0&2.198\\
\footnotemark[1]J235827.7$-$323244&25&1&&23:58:27.50&$-$32:32:44.8&2.912\\
J011424.0$-$333614&27&1&&01:14:24.01&$-$33:36:16.5&4.509\\ 
\footnotemark[1]J230815.6$-$343801&28&1&&23:08:15.73&$-$34:38:00.5&3.925&0.840\\
J235623.1$-$354119&36&1&&23:56:23.08&$-$35:41:19.5&3.095\\
\footnotemark[1]J232623.0$-$342642&37&1&&23:26:23.10&$-$34:26:44.0&2.619&0.475\\
J232900.6$-$321744&39&1&&23:29:00.81&$-$32:17:45.0&3.229&0.654\\
J013240.0$-$330907&40&1&&01:32:40.28&$-$33:09:08.0&1.971\\
J000124.9$-$354212&41&3&A&00:01:24.79&$-$35:42:11.0&4.098\\
\footnotemark[1]J000007.5$-$334060&42&3&[A+B+C]&00:00:07.45&$-$33:41:03.0&3.307\\
J005132.8$-$301848&45&3&A&00:51:32.95&$-$30:18:49.7&2.434\\
\footnotemark[1]J225250.7$-$313658&47&1&&22:52:50.76&$-$31:36:59.9&2.433&0.656\\
\footnotemark[1]J230546.3$-$331039&49&2&[A+B]&&&  &0.620\\
&&&A&23:05:46.41&$-$33:10:38.1&2.724\\
&&&B&23:05:46.58&$-$33:10:43.1&2.730\\
J013951.9$-$321446&55&1&&01:39:52.08&$-$32:14:45.5&2.656\\
J003207.7$-$303724&56&4&C&00:32:07.67&$-$30:37:34.3&2.561\\
J004853.3$-$303110&57&1&&00:48:53.38&$-$30:31:09.9&3.265\\
J005724.2$-$273122&60&1&&00:57:24.33&$-$27:31:23.3&3.261\\
J005132.0$-$302012&63&3&A&00:51:31.70&$-$30:20:20.6&2.432\\
\footnotemark[1]J223753.8$-$305828&68&1&&22:37:53.85&$-$30:58:27.9&2.719\\
J012416.0$-$310500&69&2&[A+B]&&&  \\
&&&A&01:24:16.16&$-$31:04:59.5&2.075\\
&&&B&01:24:15.87 &  $-$31:05:05.1  &2.073\\
J012853.0$-$332719&73&1&&01:28:53.07&$-$33:27:19.1&3.026\\
J005629.6$-$311206&77&2&[A+B]&00:56:29.25&$-$31.12:07.5&2.228\\
J230002.6$-$315005&80&3&[A+B]&&&&0.651\\
&&&A&23:00:02.54&$-$31:50:08.9&2.231&\\
&&&B&23:00:02.88&$-$31:50:08.0&1.968&\\
J002054.6$-$312752&81&2&[A+B]&&&&\\
&&&A&00:20:54.20&$-$31:27:57.4&3.160\\
&&&B&00:20:54.74&$-$31:27:50.8&2.588\\
J235324.7$-$331111&86&1&&23:53:24.56&$-$33:11:11.8&2.564\\
J005659.4$-$295039&90&2&A&00:56:59.28&$-$29:50:39.3&3.992\\
J234750.5$-$352931&93&1&&23:47:50.44&$-$35:29:30.2&2.400\\
J233024.1$-$325032&102&2&A&23:30:24.43&$-$32:50:32.3&3.287\\
J225324.2$-$323504&103&1&&22:53:24.24&$-$32:35:04.2&2.942&0.666\\
\footnotetext[1]{This source is observed with both ACA and the 12-m Array.}

J001802.2$-$313505&106&2&A& 00:18:02.46 & $-$31:35:05.1 &2.369\\
J014520.0$-$313835&107&1&&01:45:20.07&$-$31:38:32.5&2.553\\
J223942.4$-$333304&111&1&&22:39:42.34&$-$33:33:04.1&2.371&1.3\\
J000806.8$-$351205&117&2&A&00:08:07.20&$-$35:12:05.0&4.526\\
J012222.3$-$274456&120&2&[A+B]&&&\\
&&&A&01:22:22.44&$-$27:44:53.7&3.125\\
&&&B&01:22:22.13&$-$27:44:59.0&3.124\\
J223615.2$-$343301&121&2&A&22:36:15.31&$-$34:33:02.3&3.741\\
J003717.0$-$323307&122&2&A&00:37:16.69&$-$32:32:57.4&2.883\\
J233037.3$-$331218&123&1&&23:30:37.45&$-$33:12:16.8&2.170\\
J225339.1$-$325550&131&2&B&22:53:39.50&$-$32:55:52.3&2.197\\
J231205.2$-$295027&132&1&&23:12:05.31&$-$29:50:26.5&2.473&0.652\\
J225611.7$-$325653&135&2&A&22:56:11.79&$-$32:56:52.0&2.401&0.640\\
J011730.3$-$320719&138&2&B&01:17:30.74&$-$32:07:18.0&1.407\\
J224759.7$-$310135&141&1&&22:47:59.75&$-$31:01:35.7&2.085&0.653\\
J012335.1$-$314619&145&2&A&01:23:34.65&$-$31:46:23.6&2.730\\
J232210.9$-$333749&146&2&B&23:22:10.62&$-$33:37:58.4&2.003&0.760\\
J000330.7$-$321136&155&2&A&00:03:30.65&$-$32:11:35.1&3.077\\
J235122.0$-$332902&159&2&[A+B]&&&\\
&&&A&23:51:21.76&$-$33:29:00.4&2.236\\
&&&B&23:51:22.36&$-$33:29:08.1&2.235\\
J011014.5$-$314814&160&1&&01:10:14.46&$-$31:48:15.9&3.955\\
J000745.8$-$342014&163&3&A&00:07:46.24&$-$34:20:03.0&3.140\\
J225045.5$-$304719&168&2&A&22:50:45.48&$-$30:47:20.3&2.583&0.470\\
J011850.1$-$283642&178&4&[A+B+C]&&&\\
&&&A&01:18:50.26&$-$28:36:43.9&2.658\\
&&&B&01:18:50.09&$-$28:36:40.6&2.655\\
&&&C&01:18:49.98&$-$28:36:43.2&2.656\\
J230538.5$-$312204&182&1&&23:05:38.80&$-$31:22:05.6&2.227&0.778\\
J234955.7$-$330833&184&1&&23:49:55.66&$-$33:08:34.4&2.507\\
J225600.7$-$313232&189&1&&22:56:00.74&$-$31:32:33.0& 3.300&0.672\\
J014313.2$-$332633&200&1&&01:43:13.30&$-$33:26:33.1&2.151\\
J005506.5$-$300027&207&1&&00:55:06.51&$-$30:00:28.3&1.569\\
J225744.6$-$324231&208&2&[A+B]&&&\\
&&&A&22:57:44.59&$-$32:42:33.0&2.478\\
&&&B&22:57:44.83&$-$32:42:32.8&2.483\\
J224920.6$-$332940&209&2&A&22:49:21.04&$-$33:29:41.5&2.272&0.508\\

\end{longtable}
\twocolumn

\subsection{ACA observations}
\begin{figure}
\begin{center}
\includegraphics[width=\linewidth]{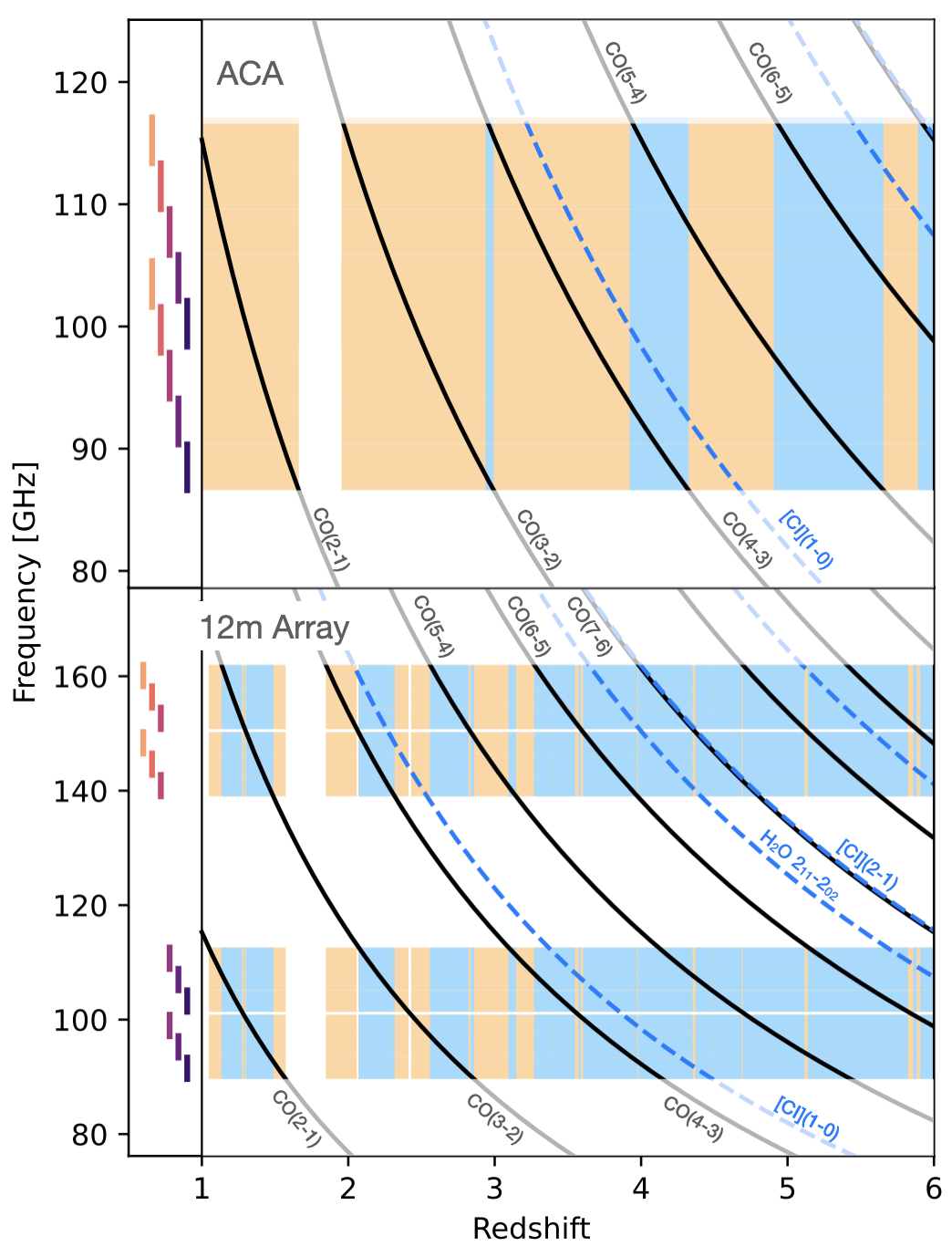}
\caption{Redshift as a function of frequency for the ACA Band 3 spectral coverage (86.6-115.7~GHz; \textit{top panel}) and ALMA 12-m\ Array (89.6-112.6 and 139-162~GHz; \textit{bottom panel}) for the $^{12}\rm CO$ (\textit{solid black lines}), \ion{C}{I} and water ({\it dashed blue lines}) emission lines. 
The \textit{blue areas} identify the redshift range where at least 2 CO emission lines fall within this frequency range, and the \textit{orange areas} show where only one CO line falls within the frequency range.  ``Redshift deserts" are shown as white areas.  The coloured bars at the left-hand side of the plots show the location of the spectral tunings that were used for the ACA and 12-m\ Array. Our 12-m\ Array tunings (Section~\ref{sec:12mobservationsdescription}) dramatically improved our ability to find robust redshifts drastically from an estimated 12 to 90 per cent using only one extra tuning.}
\label{spectralrange}
\end{center}
\end{figure}

Data were acquired with the ACA (also called the Morita Array) during ALMA Cycles 4 and 6 in programmes 2016.2.00133.S and 2018.1.00804.S (P.I.: S. Serjeant).  Some details about these observations are listed in Table~\ref{tab:imagesettings}.  The observations of each target consisted of single pointings with a series of 5 spectral tunings set up to cover the sky frequency range between 86.6 and 115.7~GHz (2.59 - 3.46 mm; Fig. \ref{spectralrange} top panel).  Each spectral tuning consisted of four spectral windows that were 2~GHz ($\sim$5500~km~s$^{-1}$) in size, with two spectral windows placed adjacent to each other in a lower sideband and two more in an upper sideband, with the sidebands separated by 8~GHz.  Not all observations with all spectral tunings were executed in Cycle 4.  For some targets, however, line emission was detected using the limited Cycle 4 data that were acquired.  For these sources, we did  not request any additional Cycle 6 observations to complete the coverage of the 86.6 to 115.7~GHz range.
Typically, several targets located close to each other in the sky were observed with the same spectral settings within one Execution Block, and thus the sources share bandpass, flux density, and phase calibrators.  The calibrators were typically quasars, although Solar System objects were used for flux calibration in some observations.

Our spectral coverage allows us to detect CO lines between the (2$-$1) and (6$-$5) transitions depending on the redshift of the sources as seen in Figure \ref{spectralrange}.  In addition, we can also potentially detect the [C{\sc I}]($^{3}\rm P_{1}-^{3}\rm P_{0})$ fine-structure line at 492~GHz for \textit{z}=3.3-5.7 sources and the [C{\sc I}]($^{3}\rm P_{2}-^{3}\rm P_{1})$ at 809~GHz for \textit{z}=7.0-9.3 sources.  Given the redshift range of the sources, other molecular lines might be expected (e.g., H$_{2}$O, HCO$^{+}$, HCN, CN, \citealp{spilker2014}), but these would not be detectable in our data given the limited observation time per source.  The CO and [C{\sc I }] lines covered in the 86.6$-$115.7~GHz range would allow us to measure spectral lines between $0.0<z<0.3$, $1.0<z<1.7$, or $2.0<z<9.6$.

The sensitivity goals were set to 3\,mJy, as measured within a velocity width of 308~km\,s$^{-1}$, since this was the expected line width of the data; this was typically matched by our observations. Our sensitivity goal was determined by the relationship between bolometric ($40-500$\microns) and CO luminosities \citep{solomon2005}, where, for the purposes of observation planning, no correction was made for differing excitation of CO lines (\cite{bothwell2013} and references therein).  For the transitions in the Band 3 window ($84-116$\,GHz), we predicted CO fluxes of $4.6-8.0$\,Jy\,km\,s$^{-1}$ for our targets, which is consistent with our CO detections of H-ATLAS lensed galaxies to date \citep[e.g.,][]{neri2019} and broadly consistent with the predictions of \cite{combes1999} once the latter is scaled to our H-ATLAS fluxes. The spectral resolution was set to 7.813 MHz, which is equivalent to a velocity bin of $20-25$\,km\,s$^{-1}$. These channels could be binned as needed to improve the detection of the line emission.

\subsection{12-m Array observations}
\label{sec:12mobservationsdescription}
Data were acquired with the ALMA 12-m Array during ALMA Cycle 7 in programme 2019.1.01477.S  (P.I.:  S. Urquhart).  Each field was observed in Bands 3 (if it had not been observed with the ACA in this band) and 4 using a single pointing with the 12-m Array in either the C43-1, C43-2, or C43-3 configurations.  These configuration yielded beams with full width at half-maxima (FWHM) of approximately 3 arcsec in Band 3 and 2~arcsec in Band 4.  Additional details of the observations are listed in Table~\ref{tab:imagesettings}.  Note that these configurations are sufficient for detecting line emission, which was the primary intention of this programme, and the configurations can also be used to resolve {\it Herschel} sources into multiple individual galaxies; however, most of the detected targets will be unresolved point sources.

The observations used six spectral tunings with three in each band, as shown in Figure~\ref{spectralrange}.  These observations produce near-continuous coverage of 23.25\,GHz bandwidth centred at 101 and 151 GHz (89.6$-$112.6 and 139$-$162~GHz).  We based the precise tunings in Bands 3 and 4 on the method detailed in \cite{Bakx2020IRAM}, where we optimised the expected number of sources with robust redshifts (x, two or more spectral lines) assuming that the proposed sources follow the existing redshift distribution of {\it Herschel}-selected galaxies (\citealt{bakx18,Bakx2020IRAM,neri2019} and references therein). We derive this optimised solution using a Monte-Carlo approach, where we generated 1000 fake redshift catalogues of 1000 redshifts, drawn from the previously-mentioned redshift distribution assuming a standard error of $\Delta z/(1+z)=0.13$ (e.g., \citealt{pearson13,ivison16}). We test all potential ALMA tuning configurations using between four and eight ALMA tunings, placed randomly across Bands 3 to 6. 

This optimisation indicated that {\it stacking} the tunings to create a continuous coverage is always favoured. The typical ALMA tuning used in studies such as \cite{weiss2013} and our earlier ACA data --- to essentially fully cover Band 3, as in the top panel of Figure~\ref{spectralrange} --- was found to be particularly good at detecting at least one spectral line (an expected 87 per cent across our sample). However, this approach is rather inefficient given the large overlap between the upper sidebands of the lower-frequency tunings and lower sidebands of the higher-frequency tunings. Similarly, this set-up only results in a robust multi-line detection for galaxies beyond redshift 3.5 (an expected $\sim$12 per cent of our sample). Instead, two sets of three tunings in Bands 3 and 4 significantly increased the number of sources with robust redshifts to an expected 65 per cent, and it diminished the size of the {\it redshift desert} around redshift 2. An added benefit from this larger per centage of robust redshift detections, is that we can exclude redshift solutions where we would have detected more than one line. In fact, we can identify the spectroscopic redshifts of sources robustly with just a single line, by excluding redshift solutions that would have resulted in multi-line detections. In other words, if we would have detected two spectral lines within our covered bandwidth for every redshift solution except one, we can identify the redshift robustly from a single bright detected spectral line. This method is explained in more detail in \cite{Bakx2020IRAM}, and a dedicated discussion of this method will be presented in Bakx et al. in prep..  In total, this statistical exercise raised the probability for detecting robust redshifts (both by multiple spectral line, and by inference) from 12 per cent to 90 per cent while only requiring 20 per cent extra observation time for the additional band 4 tuning. Based on the results of the ACA campaign described above, we required an RMS of 0.8\,mJy for a 300\,km\,s$^{-1}$ line; this was typically matched by our observations.

\subsection{Data processing}

The 12-m Array data were pipeline-calibrated with the {\sc Common Astronomy Software Applications} ({\sc CASA}) package version 5.6.1 \citep{mcmullin2007}, while the ACA data were manually calibrated with the same version of CASA.  In both cases, the first steps of the calibration of the visibility data included amplitude corrections based on the system temperature and antenna position corrections.  Phase corrections based on water vapour radiometer measurements were also applied at this point to the 12-m Array data only.  After this, we flagged shadowed antennas and channels with low sensitivities at the edges of the spectral windows.  In the ACA data, we also visually inspected the data and flagged any data where the amplitude gains were outliers, where the amplitudes varied irregularly across the spectral window, or where the phases show jumps between observations or unusually high scatter.  For both the ACA and 12-m Array data, we then calibrated the amplitudes and phases both as a function of channel and as a function of time.  The uncertainty in the flux calibration is 5 per cent \citep{remijan2019}.

Imaging was done with the {\sc tclean} command within {\sc CASA} version 5.6.1.  Slightly different settings were used for the Band 3 ACA data, the Band 3 12-m Array data, and the Band 4 12-m Array data, and these different settings are listed in Table~\ref{tab:imagesettings}.  All images were created using natural weighting, the standard gridder, and the Hogbom deconvolver.  The pixel scale was set to so that the full-width at half-maximum (FWHM) of the beam was sampled by at least three pixels, and the image size was set to cover the area over which the primary beam is $>$0.20$\times$ the peak value.  Spectral lines were initially identified in the image cube without continuum subtraction.  Once spectral lines were found, the continuum was subtracted from the visibility data, and the data were re-imaged with manual adjustments made to the channel width in the final image cubes to optimize the balance between the sampling of the spectral line emission and the signal-to-noise ratio (as measured in both the individual image slices and the extracted spectrum).

\begin{table*}
\caption{ALMA image characteristics.}
\label{tab:imagesettings}
\begin{tabular}{lccccccc}
\hline
Array &
  Band &
  Central &
  Total &
  Typical &
  Pixel  &
  \multicolumn{2}{c}{Image size} \\
&
  &
  frequency &
  bandwidth &
  beam FWHM &
  scale &
  [pixels] &
  [arcsec] \\
&
  &
  [GHz] &
  [GHz] &
  [arcsec] &
  [arcsec] &
  &
  \\
\hline
ACA &
  3 &
  101 &
  29.1 &
  17 $\times$ 10 &
  2.0 &
  100 $\times$ 100 &
  200 $\times$ 200 \\
12-m Array &
  3 &
  101 &
  23.25 &
  3.6 $\times$ 2.7 &
  0.5 &
  240 $\times$ 240 &
  120 $\times$ 120 \\
12-m Array&
  4 &
  151 &
  23.25 &
  2.2 $\times$ 1.8 &
  0.3 &
  240 $\times$ 240 &
  72 $\times$ 72\\
\hline
\end{tabular}
\raggedright \justify \vspace{-0.2cm}
\end{table*}

\section{Spectroscopic Redshifts}\label{sec:results}

\subsection{Line extraction}

Line measurements were made using aperture photometry within the image cubes.  Lines were first identified by visual inspection of the data independently by several of the authors, and then confirmed through the following process. Circular apertures were centered on the peaks of the corresponding continuum emission, and the radii of the apertures were manually adjusted for each source in each image to include as much line emission as possible while still measuring that emission at higher than the 5$\sigma$ level.  Similarly, we manually selected frequency channels that measured as much of the line flux as possible without diluting the signal with background noise so much that the measurements fall below 5$\sigma$.  Note that the continuum is detected at a higher S/N than the line emission.  If we matched the apertures used to measure the continuum and line emission, we would either need to choose large apertures that included all of the detectable continuum emission but also included extra noisy pixels for the line emission or small apertures that optimize the line detections but do not include all of the continuum emission.

\subsection{Redshift determination}
With the precise frequencies of the spectral lines in hand, we calculate all potential redshift solutions for each line, and use the method described in \cite{Bakx2020IRAM} to provide only robust redshifts. This method accounts for any sources that could be influenced by the redshift degeneracy that can affect the linear CO-ladder\footnote{As an example of this degeneracy, the observation of \textit{J}$_{\rm up} = 2$ and 4 CO-line transitions of a $z = 2$ galaxy (76.7 and 153~GHz, respectively) could also be interpreted as the \textit{J}$_{\rm up} = 3$ and 6 CO-line transitions of a $z \approx 3.5$ galaxy.}.
In total, we find redshifts for 59 sources using multiple spectral lines that point to an unambiguous redshift solution.
Meanwhile, this method also provides additional information. For thirteen sources, we find bright emission from only a single line. In these specific cases (HerBS-22, -39, -40, -60, -73, -80B, -81A, -103, -122A, -146B, -155 and -207, -208B), we are able to exclude all other redshift solutions.  The exclusion of redshift solutions requires us to be confident that lines are indeed non-existent. Since adjacent CO spectral lines typically have similar integrated line fluxes, we can only exclude redshift solutions for galaxies with strong line detections in CO lines.

We note that the uncertainty of the spectroscopic redshifts is less than 0.001, however for clarity, we show only three trailing digits in Table \ref{tab:data_table}.  For the SGP field, data were taken from H-ATLAS SGP Data Release 2 Catalogue version 1.4 \citep{smith2017,maddox2018,furlanetto2018}.  There are fields where we have detected multiple sources.  In these cases, we have labelled them alphabetically with decreasing brightness and quoted redshifts for the sources where we were able to robustly detect them.  Details on the continuum measurements, including information on additional sources detected only in continuum emission, will be given in Bendo et al. (in prep).  In cases where our lowest angular resolution does not resolve the source into separate components, we only provide the redshift of the \textit{system}, denoted by straight brackets (e.g., HerBS-21 [A+B] at $z_{\rm spec} = 3.323$).

\begin{figure*}
\begin{center}
\includegraphics[width=\textwidth]{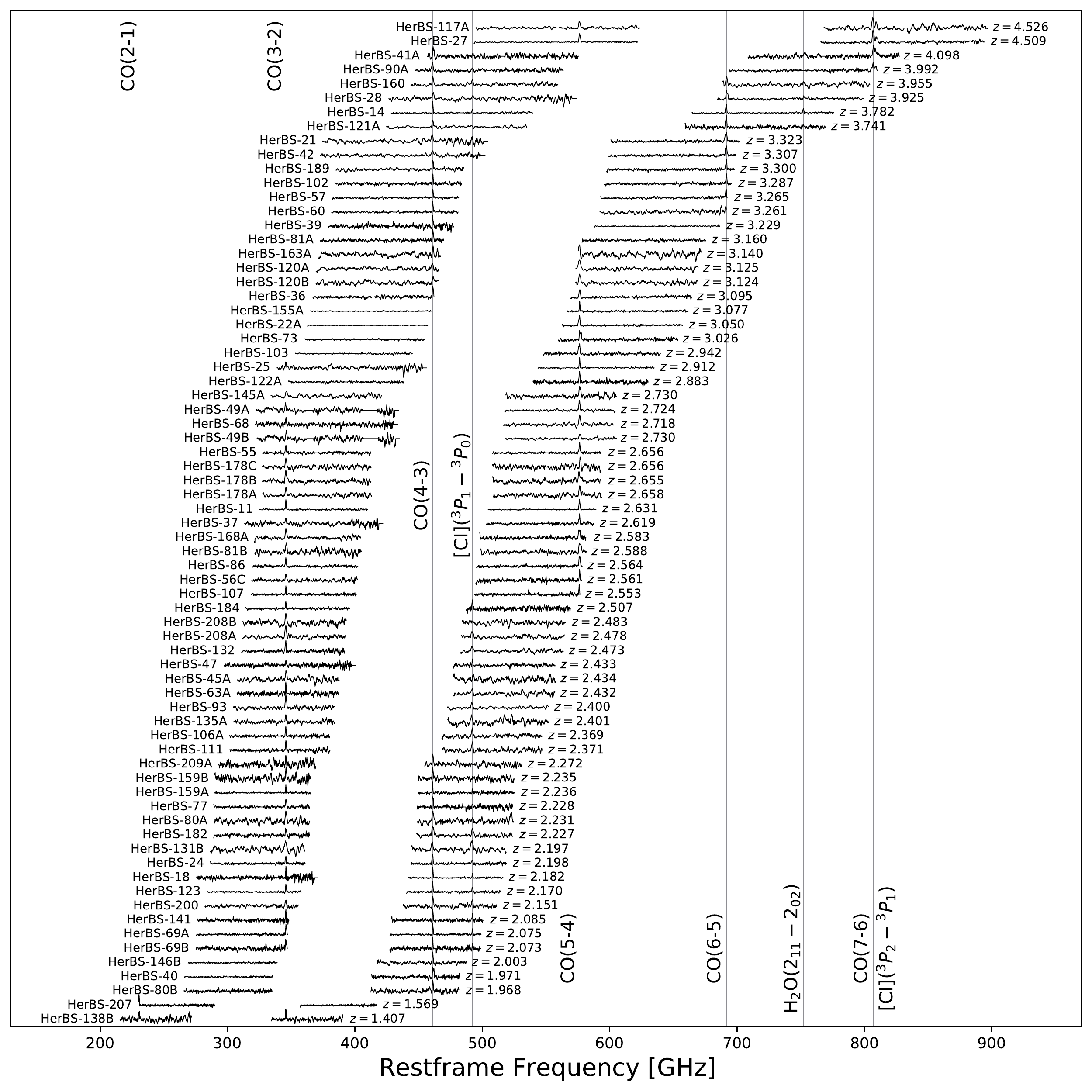}
\caption{ALMA spectra of the 71 galaxies with spectroscopic redshifts reported in this paper shown at their rest-frame frequency, offset vertically for clarity. Vertical lines indicate the transitions of CO, H$_2$O and \ion{C}{I} lines. The sources are ordered by redshift and the flux scaling for each spectrum is arbitrary. The optimised 3- and 2-mm band tunings resulted in a successful redshift identification for the majority of sources.}
\label{combinedspectra}
\end{center}
\end{figure*}

In total, across our 71 galaxies with robust redshifts (associated with 62 {\it Herschel} sources, Table 1), we find the following lines robustly (for upper limits we refer to the upcoming paper by Hagimoto et al.). We primarily detect CO(3$-$2), (4$-$3) and (5$-$4) emission  (38, 36 and 28 sources, respectively), with only a handful of sources with CO(2-1), CO(6-5) and CO(7-6) emission (two, nine and four, respectively).  For 21 galaxies we find CI($^{3}\rm P_{1}-^{3}\rm P_{0}$) emission, for one galaxy CI($^{3}\rm P_{2}-^{3}\rm P_{1}$) emission, and one galaxy shows H$_2$O 2$_{11}$-2$_{02}$ emission.  We detail the line fluxes in Table~\ref{tab:lines}, and the lines can be seen graphically in Figure~\ref{combinedspectra}.  Here we note that these individually-resolved components are likely to be individual galaxies (e.g., \citealt{Hayward2013}).  However, it is possible that a small fraction are multiple images of the same galaxy system lensed by foreground cluster lenses. The exact nature of the individual components (among which are protocluster cores; e.g., \citealt{Oteo2018}) will be discussed more in upcoming papers. Multiple images in galaxy-galaxy gravitational lensing systems would not be resolved in our ALMA data, since for source redshifts $\gg$ lens redshifts, the critical radius in a Singular Isothermal Sphere lens asymptotes to $\sim1.5^{\prime\prime}\times(\sigma_{\rm v}/230$\,km\,s$^{-1})^2$ where $\sigma_{\rm v}$ is the lens velocity dispersion.

\label{sec:spectroredshifts}

\subsection{Sources without robust redshifts}
\label{sec:singlelines}
We were unable to identify the redshifts for 23 \textit{Herschel} sources with the ALMA 12m-Array data.  Only for five of these sources did we not detect any line emission, indicative that our true redshift desert is relatively small.  For seven of these targets, we did not deem the main line bright enough to use the exclusion method to remove any of the adjacent redshift options (two of which had ACA Band 3 observations instead of from the ALMA 12m-Array).  Nine targets have suggestive secondary lines, although none of them are above the 5$\sigma$ threshold to result in a redshift detection, and finally, seven sources have only a single line observed, with no ability to exclude any nearby redshift solutions, and thus remain ambiguous in their redshift solution.  The objects without redshifts do not otherwise appear to be atypical of our sample, as can be seen in Fig.\,\ref{fig:colour_selection} and Bendo et al. in prep. We are currently underway with, and planning future follow-up observations to reveal the redshifts of these remaining 23 sources, and will provide the complete catalogue of redshifts in a future work. 

\begin{figure}
\includegraphics[scale=0.6]{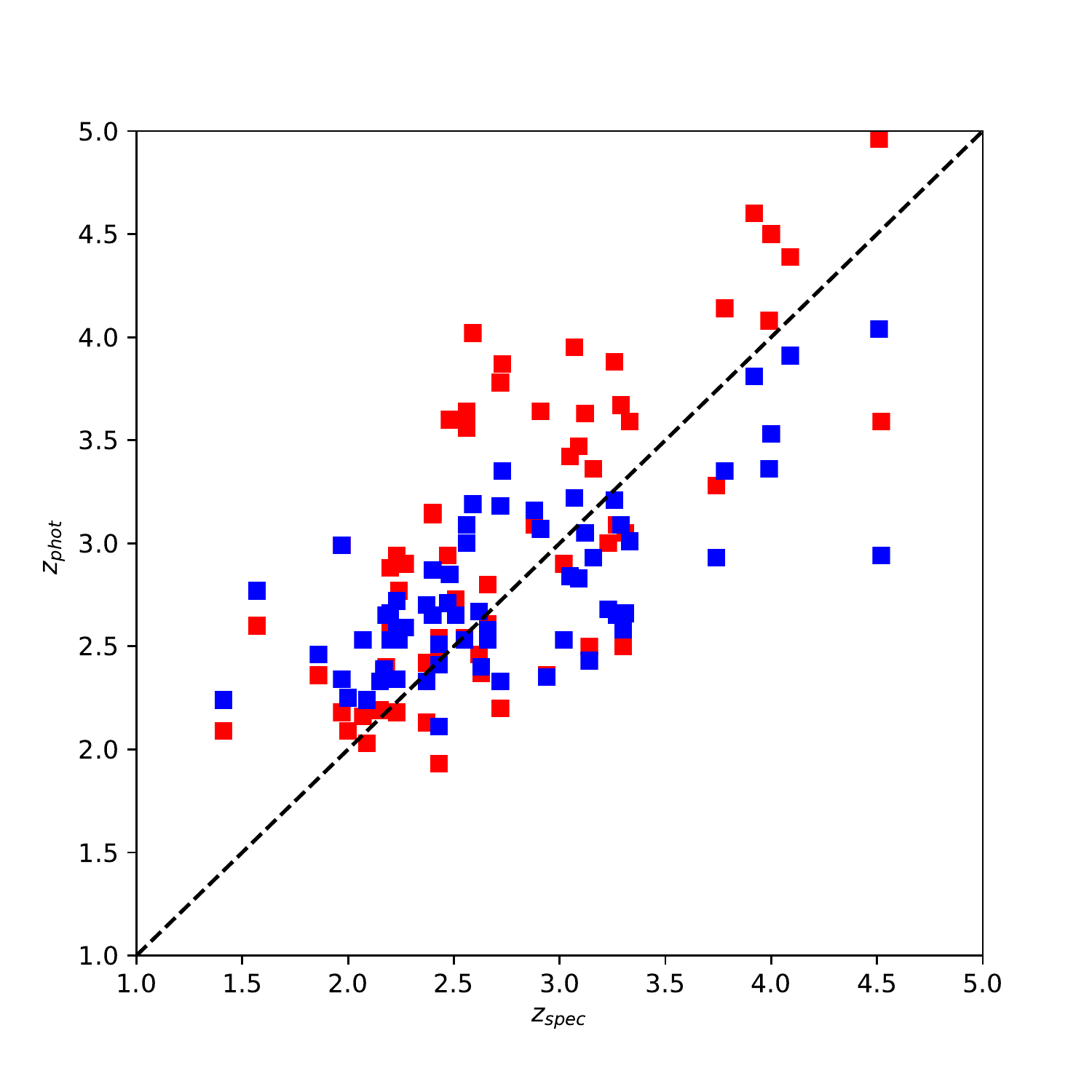}
\caption{Comparison of the spectroscopic redshifts presented here with the photometric redshifts from \citet{bakx18} shown in red and from \citet{ivison16} shown in blue.  Our results are consistent with the spread seen in, e.g., \citet{ivison16}, \citet{bakx18}, \citet{pearson13}. Note that this diagram is necessarily restricted to objects with spectroscopic redshifts.} 
\label{fig:spec_v_phot}
\end{figure}

\subsection{Spectroscopic versus photometric redshifts}

Fig.~\ref{fig:spec_v_phot} presents a comparison of the photometric redshifts from previous catalogues in \cite{Bakx2020Erratum} which use an SED fit and the method described in \cite{ivison16} to the spectroscopic measurements presented here, showing a dispersion similar to the ones seem in \cite{ivison16} and \cite{bakx18}.  For \textit{Herschel} sources with multiple components at the same redshift (HerBS 49, 69, 120, 159, 178 and 208), we show the weighted average of the spectroscopic redshift against the photometeric redshift estimate. The two sources with multiple components at different redshifts are excluded in this figure.  We find similar uncertainties on the submm photometric redshifts as previously reported in \cite{pearson13,ivison16,bakx18,jin2019}.
The average difference between $z_{spec}$ and $z_{phot}$, $\Delta z$, for the two samples are $\Delta z_{\rm Bakx}$= 0.469 and $\Delta z_{\rm Ivison}$= 0.388. The standard deviations are $\sigma(\Delta z_{\rm Bakx} / (1+z_{\rm spec}))=0.14$ and
$\sigma(\Delta z_{\rm Ivison} / (1+z_{\rm spec}))=0.13$. Paper II (Bendo et al. in prep.) will present a more exhaustive analysis of the continuum spectral energy distributions and the photometric redshifts.  

\subsection{Comparison to foreground lens redshifts}
\label{subsec:foreground}

Foreground lens redshifts and morphologies are the subject of multiple ongoing optical and near infrared imaging and spectroscopy campaigns with \textit{HST} (e.g., \citealt{berta2021}, see also Borsato et al. in prep.), \textit{Spitzer} and large ground-based telescopes (e.g., VLT and Keck).  Submm-selected strong lens candidates are found purely on the basis of the magnification bias, and unlike optically-selected or radio-selected lenses, the selection is independent of lensing morphology, lensing galaxy properties, or the presence of emission lines in the background source.  This relatively clean magnification-based lensing detection, together with the negative submm K-correction that permits detection of background sources to $z=5$ and beyond, means that we are sensitive to foreground lenses out to $z\sim$2 and therefore, these follow-up programmes can be used to probe the evolution of stellar and halo mass distributions.  For sources where the data currently exist, we compare our spectroscopic lensed redshift values with those of proposed foreground lensing galaxies taken from \cite{bakx2020}, derived from the VISTA Kilo-degree Infrared Galaxy (VIKING) survey \citep{edge2013}, a survey in \textit{zYJHK}$_{\rm S}$ to sub-arcsecond resolution.  This survey overlaps with both the equatorial GAMA fields and the Southern Galactic Pole (SGP) fields and thus covers a number of the H-ATLAS sources.  At the time of writing, however, no VIKING catalogue has been published over the whole SGP and GAMA fields.  For the 98 HerBS sources in their sample, \cite{bakx2020} found probable lenses for 56 and showed that, within 10 arcseconds, 82 per cent of the HerBS sources have associated foreground VIKING galaxies.  Table~\ref{tab:data_table} gives the probable lens redshifts applicable here and, as expected, they all have lower photometric redshifts than the sources we present.

\section{Discussion}\label{sec:discussion}
\subsection{Redshift and CO line width distributions}\label{sec:redshifts}

\begin{figure}
\includegraphics[width=\linewidth]{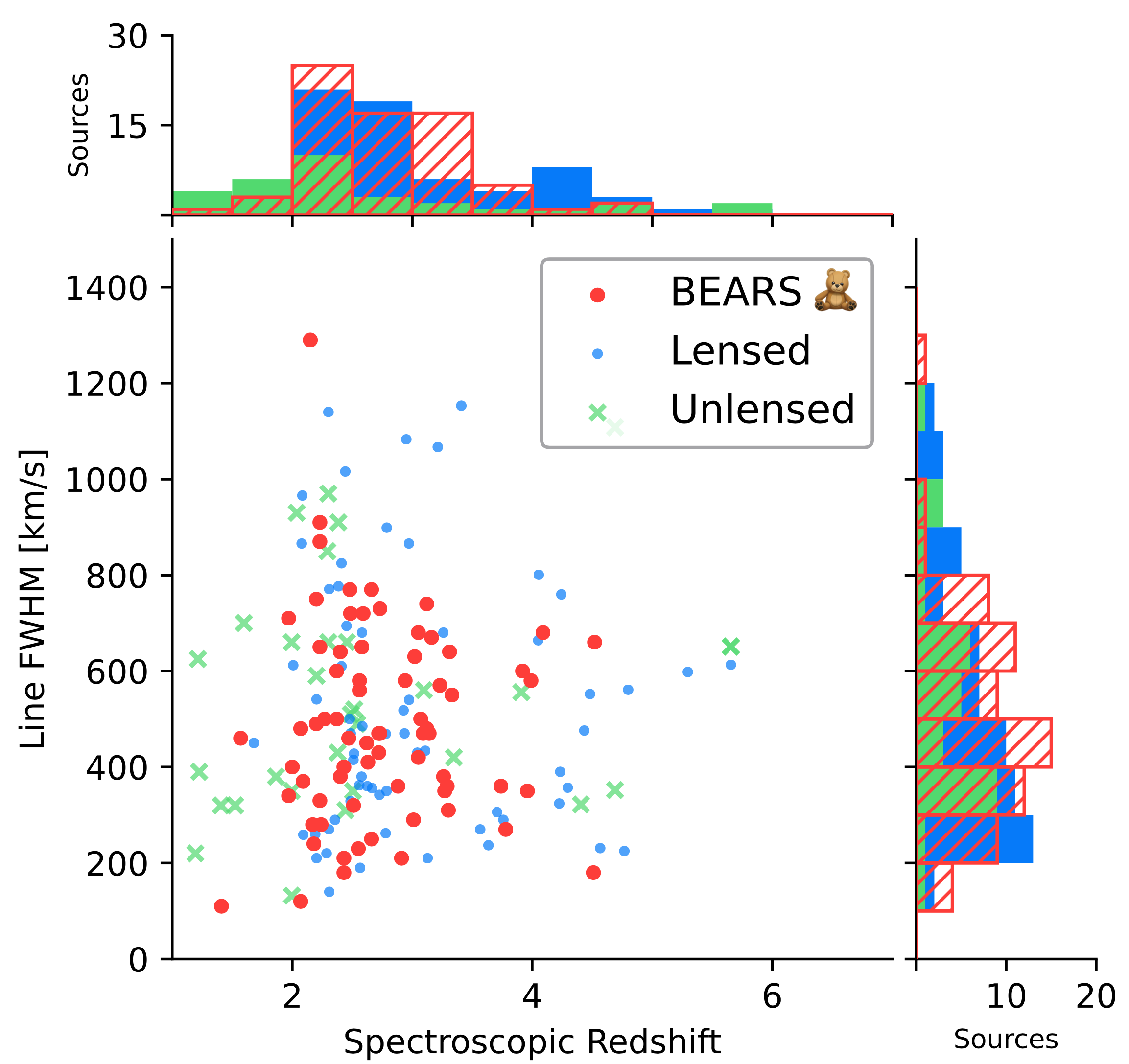}
\caption{Distribution of the spectroscopic redshifts and linewidths of the 71 BEARS galaxies presented in this work, compared to the samples described in the text.} 
\label{fig:zdist}
\end{figure}

We derived CO brightness estimates via equation~3 of \cite{solomon2005}:
\begin{multline}\label{eqn:solomon_eqn3}
    \frac{L^{\prime}_{\rm CO}}{\rm K\,km\,s^{-1}\,pc^{2}}  =  
  3.25\times10^7 \times \\
  \frac{S_{\rm CO} \Delta v}{\rm Jy\,km\,s^{-1}} 
    \left( \frac{\nu_{\rm obs}}{\rm GHz} \right)^{-2}
    \left( \frac{D_{\rm L}(z)}{\rm Mpc}\right )^{2 }
    (1+z)^{-3},
\end{multline}
\noindent where $D_{\rm L}$  is the luminosity distance to redshift $z$, $\Delta\nu$ is the linewidth, $\nu_{\rm obs}$ is the observed frequency, and $S_{\rm CO}$ is the observed line flux.  Figure \ref{fig:zdist} shows the distribution of redshifts and line widths (FWHM) for the 71 galaxies with robust spectroscopic redshifts from this work. We compare them against two samples, roughly divided into a lensed and unlensed sample. The lensed sample is compiled from \citet{neri2019}, \citet{harris2012}, \citet{yang2017}, \citet{aravena2016} and some well known individual sources.  These are the lensed sources known as IRAS\,FSC\,10214+4724 \citep[e.g.,][]{rowanrobinson91,serjeant95,eisenhardt1996}, the Cosmic Eyelash \citep{swinbank2010}, the Cloverleaf Quasar \citep[e.g.,][]{cloverleaf_discovery,cloverleaf_co,cloverleaf_hcn}, APM\,08279+5255 \citep[e.g.,][]{apm08279_discovery} and the Cosmic Eyebrow \citep[e.g.,][]{dannerbauer2019}.  The unlensed sample comes from \citet{bothwell2013}, \citet{harris2010} and several well known individual unlensed sources from literature.  These are the sources known as BR\,1202$-$0725 (N and S) and BRI\,1335$-$0417 \citep{carilli2013,dannerbauer2019}, along with two binary hyper-luminous galaxies: HATLAS\,J084933+021443  \citep[$z$=2.41,][]{ivison2013} and HXMM01 \citep[$z$=2.308,][]{fu2013}.  

The redshift distribution of our sample has a mean redshift of 2.75 and median redshift of 2.61, with all sources lying within the redshift range $1.41< z< 4.52$.  
For 67 of the sources, the exceptions being HerBS-40 ($z=1.971$), HerBS-80B ($z=1.968$), HerBS-138B ($z=1.41$) and HerBS-207 ($z=1.57$), we find redshifts above 2, which we would expect due to the photometric redshift pre-selection.  The mean redshifts of these lensed and unlensed comparison samples are 3.04 and 2.65, respectively.

The mean line width (FWHM) distribution is also shown in Fig~\ref{fig:zdist}. We find a median width of 475~km~s$^{-1}$ and a mean of 494~km~s$^{-1}$, which lies below both the values of the lensed and unlensed samples reported which are 524~km~s$^{-1}$ and 577~km~s$^{-1}$ respectively, the median values are 496~km~s$^{-1}$ (lensed) and 531~km~s$^{-1}$ (unlensed). As we will discuss later, this is consistent with their lensing nature and we will discuss the physical interpretation of this, including their complex dynamics in subsequent sections.

\subsection{The  $L^{\prime}_{\rm CO}$ and $\Delta V$ relation}\label{sec:lprime}

\citet{bothwell2013} noted that the CO luminosities of submillimetre galaxies correlate with their line widths, which they interpreted as baryon-dominated gas dynamics. This relation extends a similar trend seen in molecular clouds \citep{Bolatto2008}. Lensed submillimetre galaxies occupy a different region of this relation, as noted by \citet{harris2012}, due to magnification effects predominantly affecting the luminosities and not line widths. 

\begin{figure*}
\includegraphics[scale=0.8]{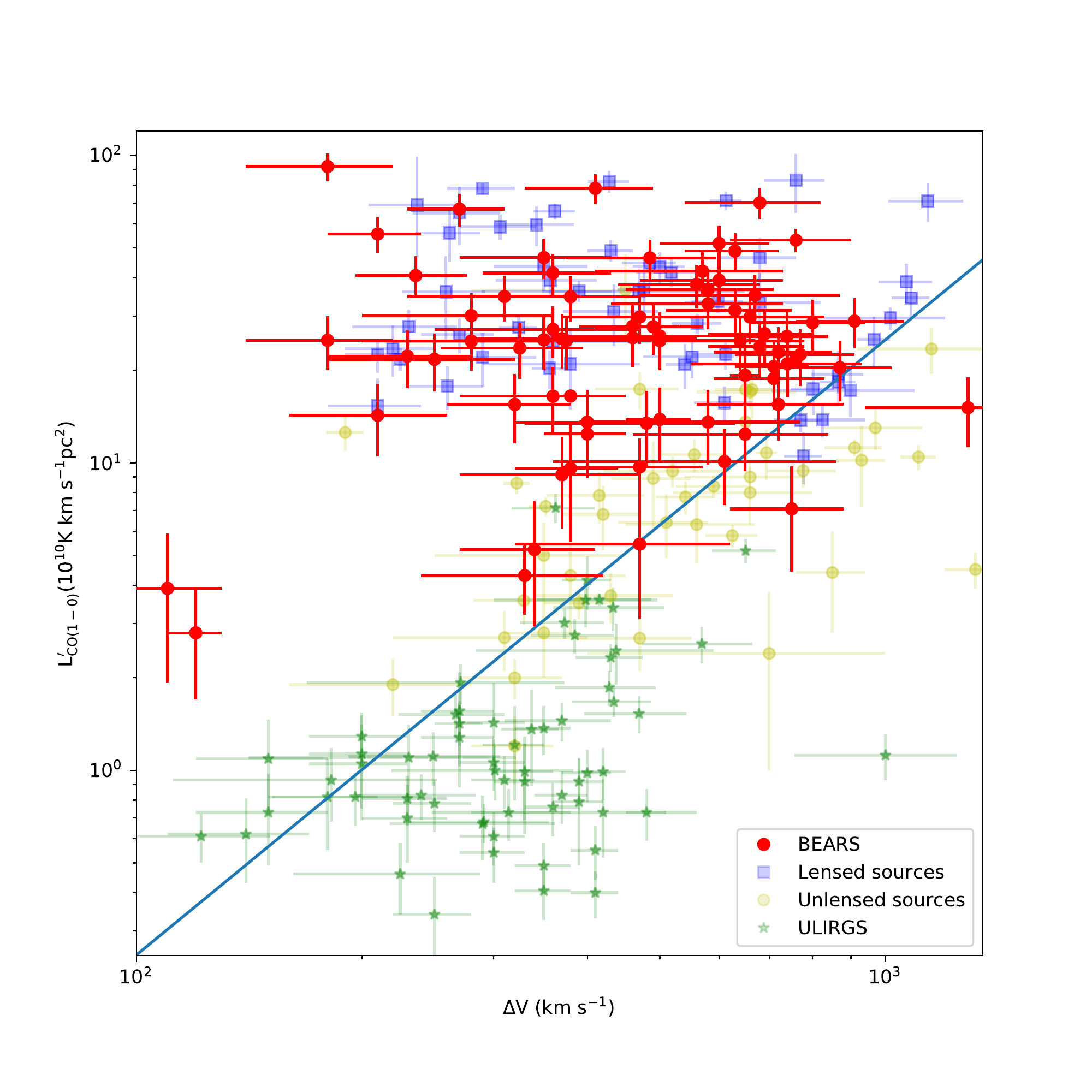}

\caption{Apparent CO luminosity ($L^{\prime}_{\rm CO(1-0)}$) versus linewidth  ($\Delta V$) for the sources identified in this work, along with sources from literature.  Shown are both high-redshift lensed and unlensed samples (see Section \ref{sec:redshifts} for more details) and a sample of ULIRGS (see Section \ref{sec:lprime} for more details). A trend line for the unlensed population is also shown.}

\label{fig:L_v_V}
\end{figure*}

The $^{12}$CO (1-0) luminosities of our sample were calculated following the standard relation of \cite{solomon2005} (Eq.~\ref{eqn:solomon_eqn3}).  The results are presented in Fig. \ref{fig:L_v_V} and show the relationship between apparent CO luminosities and the FWHM of the CO emission line, $\Delta V$ including data from a number of other studies.  We use the previously defined ``lensed" and ``unlensed" samples (Section 4.1) and create an additional sample of ULIRGS.  This is composed of data from \cite{solomon1997}, \cite{combes2011} and \cite{combes2013}. The values of the CO luminosities used in this paper are explicitly for CO(1$-$0), $L^{\prime}_{\rm CO(1-0)}$, in order to make a direct comparison with the majority of quoted literature values, a correction for excitation was applied using the median brightness temperature ratios for SMGs in Table 4 of \cite{bothwell2013}, and accounting for errors.  We calculate the CO(1$-$0) luminosity from the lowest CO transition available. For a minority of cases, where sources in the literature were not given as $^{12}\rm CO(1-0)$, the appropriate corrections (as already described) for excitation were applied.  It should be noted that none of these (likely) gravitationally lensed sources are corrected for the effects of lensing magnification.

Including our sources in this diagram does not alter the observed trend reported in \cite{neri2019} and also seen in, for example, \cite{harris2012} whereby there is a clear distinction between sources that are strongly lensed and those that are unlensed.

\subsection{ALMA as an efficient redshift hunter}
Our Bands 3 and 4 observations with ALMA identified the redshifts for 71 galaxies, based on the initial positions of 62 \textit{Herschel} sources (Table 1). We did not find robust redshifts for 23 \textit{Herschel} sources, although two of those only had Band 3 data from ACA. Using our bespoke tuning set-up, we have a success-rate of 75 per cent ($= 62/[62+23 - 2]$) which is lower than the initial 90 per cent promised by our simulations (see Section \ref{sec:12mobservationsdescription}). As discussed in Section \ref{sec:singlelines}, only seven sources have single lines detected without the ability to exclude any nearby redshift solutions. Deeper observations promise to improve this fraction to 92 per cent ($=[85 - 7]/[85]$), given that we failed to use our exclusion method on seven sources, and nine sources had low-significance line features that were below our line detection criteria. The low-significance line features may be partially due to the large multiplicity found for many of our sources (see Bendo et al.\ in prep. for more details).
Here, we also note, however, that this 92 per cent estimate might have been optimistic, since: (1) two sources were observed with the full five-tuning Band 3 coverage in ACA before a three-tuning follow-up in Band 4 using the 12-m\ Array; and (2) some of the nine sources with faint line emission might have redshifts that still prove ambiguous even with deeper observations, because sometimes the limiting factor is spectral coverage rather than signal-to-noise. 

\subsection{The effects of differential magnification on luminosities}
\label{sec:diff}

Could differential magnification affect the observed distribution of magnification factors evident in Fig.\,\ref{fig:L_v_V}?
Whilst gravitational lensing is a purely geometrical effect, and as such is independent of wavelength, the degree of magnification will vary depending upon the line of sight.  For example, an extended background source may exhibit intrinsic colour gradients, and the magnification may vary across the resolved background source. Therefore a lensed source and an unlensed source may have differing colours when averaged over the system. This is ``differential magnification", potentially significantly affecting broad-band photometry and crucial emission line diagnostics \citep{serjeant2012}.  It is often assumed that this effect can be neglected, but here we would like to justify why this is an acceptable assumption to make in this paper.

\begin{figure}
\includegraphics[scale=0.6]{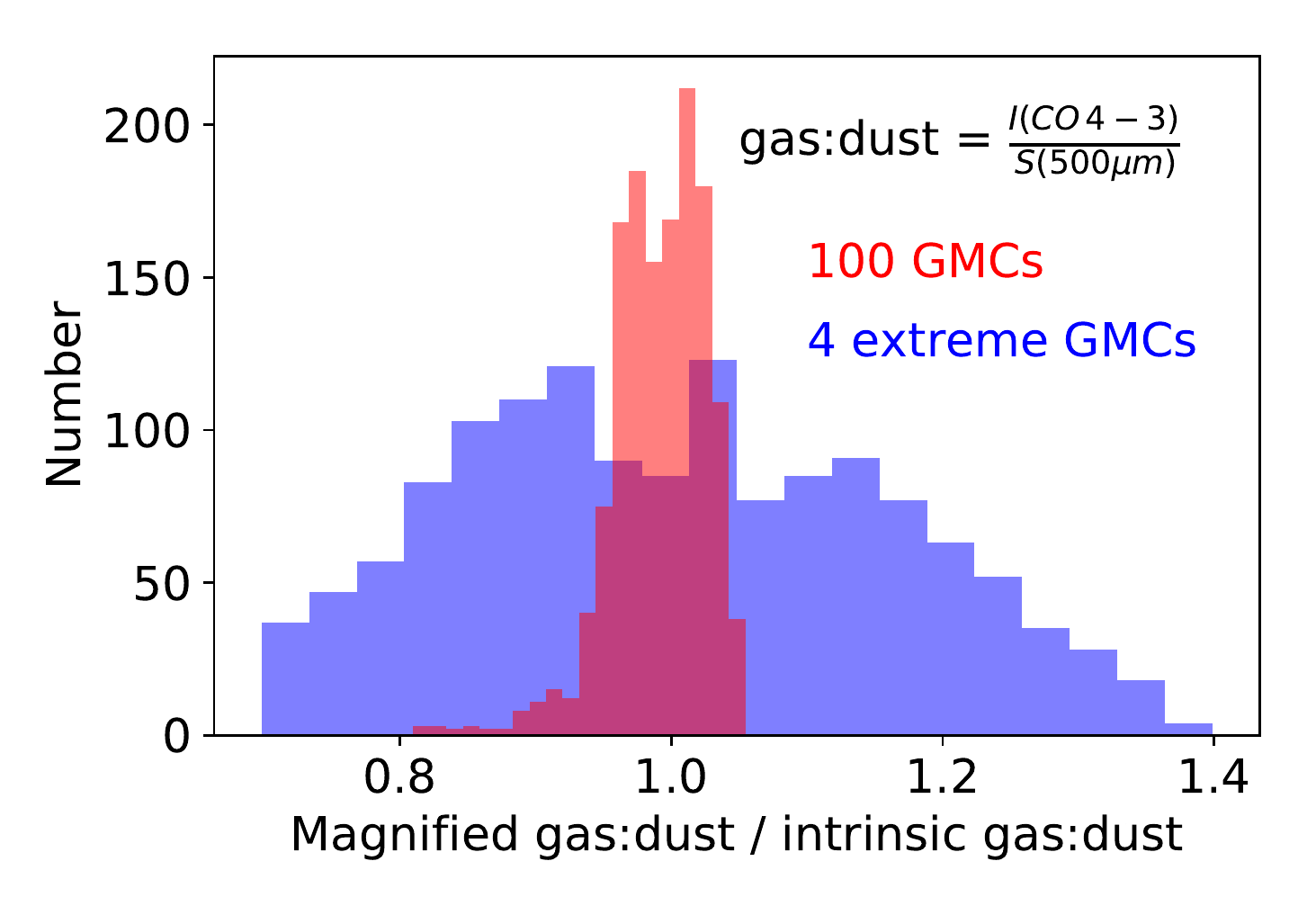}
\caption{Simulation of differential magnification effects on a submm galaxy at $z=2.286$ lensed by a $z=0.9$ foreground galaxy. The blue histogram shows random realisations assuming the star formation is confined to only four giant molecular clouds as claimed originally for the Cosmic Eyelash by \citet{swinbank2010}, while the red histogram with narrower bins shows the distribution of random lensing realisations with star formation distributed over 100 identical molecular clouds, as suggested by the smooth continuum observed by \citet{ivison2020}. Note the much narrower distribution when the star formation is spread more uniformly throughout the system (red), and the wider distribution when the star formation is concentrated into a very small number of clumps (blue).}
\label{fig:diff}
\end{figure}

We have repeated the methodology of \cite{serjeant2012} (to which the reader is referred for more details) to run a simulation of a $z=2.3$ background galaxy with a foreground gravitational lens at $z=0.9$. The background source structure was modelled on the original \cite{swinbank2010} observations of the Cosmic Eyelash lens system, with four giant star-forming clumps, selected as an ostensibly typical example of the submm galaxy population.  This is shown in Fig. \ref{fig:diff}. The $1\sigma$ spread in differential magnification effects are at around the 20 per cent level. 

However, since the \cite{swinbank2010} Submillimeter Array (SMA) observations of the Cosmic Eyelash, \cite{ivison2020} using ALMA found a much smoother dust continuum and suggested that the SMA structures were artefacts of using {\sc CLEAN} to image low signal-to-noise data, an issue that did not affect the images created from the ALMA data. The\, $\sim\!\pm20\%$ systematic in Fig.\,\ref{fig:diff} is therefore the most pessimistic case. In order to quantify the effect of more smoothly distributing the star formation, we performed a new set of simulations with star formation contained in 100 giant molecular clouds, rather than four, again following the methodology of \cite{serjeant2012}.  This is shown as the red curve in Fig.\,\ref{fig:diff}, and is a $\pm3.5\%$ effect. The differential magnification effects are larger when the CO emission is more concentrated, because it is easier for a large proportion of the CO flux to be close to a caustic curve. 

In summary, the worst case is that the differential magnification effects are comparable to or smaller than the random errors, while in more realistic up-to-date models the effects in this case are negligible. 
We conclude that we can reasonably neglect differential magnification effects for the CO lines in this paper.

\subsection{Magnification distribution}
The vertical offset of lensed sources in Fig.\,\ref{fig:L_v_V} is due to gravitational magnification, which in turn depends only on the lensing geometry and is therefore formally independent of velocity width $\Delta V$. This figure can therefore be used to make magnification estimates. The magnifications from the offset from our fit to the unlensed samples are listed in Table \ref{tab:MagnificationFactors}. 

The underlying magnification probability distribution generically tends to follow a $\Pr(\mu,z)\mathrm{d}\mu=a(z)\mu^{-3}\mathrm{d}\mu$ power law \citep{blain1996,Serjeant2014}, but this is then convolved with the luminosity function, so the resulting magnification distribution may not necessarily be representable by such a simple power-law function. Therefore, a lensing population model is required. 

\begin{figure}
\includegraphics[scale=0.6]{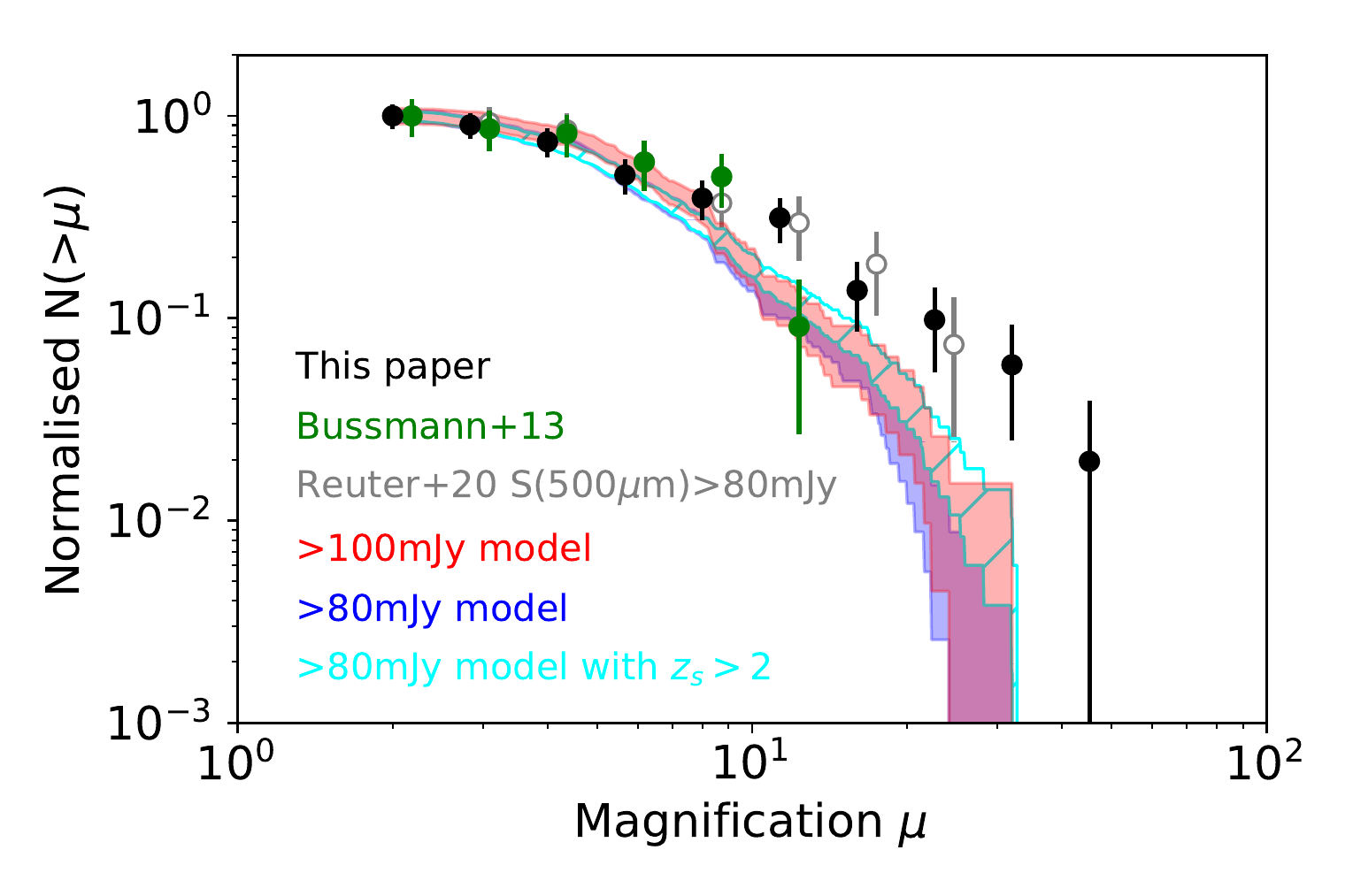}
\caption{Observed cumulative magnification distribution of lensed submm galaxies (magnifications $\mu\geq2$) with redshifts from this paper, compared to the LensPop submm population discussed in the text with a flux limit of $>100$\,mJy (red) and $>80$\,mJy (blue), and with a HERBS-like additional selection of source redshift $z_{\mathrm{source}}>2$ (cyan hatched). The shaded areas represent the Poisson uncertainties in the numbers from the model realisations. Note the good agreement between data and models over most of the magnification range, but the hint of an excess at large 
magnifications. Also shown is the magnification distribution from the 500\microns-selected sample of \citet{bussmann13}, and the South Pole Telescope targets with $\geq80$\,mJy at 500\microns\ and with modelled magnifications \citep{reuter20}, though note that the latter data points have flux limits at a total of three wavelengths so are not straightforwardly comparable.}
\label{fig:lenspop}
\end{figure}

The model used as the basis for the work presented in this section was described in \citet{COLLETT2015} with its source code available as open source software\footnote{\tt https://github.com/tcollett/LensPop}, and was originally designed to predict strong gravitational lensing of optical galaxies. As described in that paper, the code was verified against observations of lenses by the galaxy-scale search of the Canada-France-Hawaii Telescope Legacy Survey \citep{More2012,More2016,Gavazzi2014}.
The paper predicted lenses discoverable by the Wide Field Survey of \textit{Euclid} \citep{Laureijs2011}, the Dark Energy Survey \citep{Treu2018} and the Rubin Observatory \citep{Ivezic2008}. The model was used in \citet{Weiner2019} to predict $122$ strong galaxy-galaxy lenses in COSMOS, in broad agreement with the 23 highly likely lenses and 159 lens candidates in that field \citep{faure2008,jackson2008}, to make predictions for the Euclid Deep Fields and to examine dependencies on cosmological parameters. A further prediction of $\geq$17\,000 lenses discoverable by the Nancy Roman Space Telescope \citep{Green2012} was presented in \citet{Weiner2020}.

The model assumes strong lensing by elliptical galaxies, modelled as singular isothermal ellipsoids (SIEs). A population of foreground SIEs is generated with five key parameters: redshift, stellar velocity dispersion, flattening, effective radius and absolute magnitude.  The lensing cross-section of the foreground population is then projected onto a selected background source population to generate an idealised set of lens systems (deflector + source). Finally, the model applies criteria based on the observing parameters of the survey being considered in order to discover the final set of strong lenses detectable by that survey.

To adapt the model for use in predicting lensing of submillimetre galaxies, we have made two major changes: firstly, a source catalogue of submillimetre galaxies was needed to replace the source catalogue of optical galaxies originally used; secondly, the criteria used to detect the lenses were altered.
A mock submillimetre catalogue was created, based on the study by \citet{Cai2013} that allows for an estimate of number counts as a function of the (unlensed) flux at 500\microns\ and redshift. For simplicity, we have used only submillimetre galaxies at $z > 1$, and require that the background source lies within the Einstein radius in order for strong lensing to be possible. The LensPop model also requires the source galaxy angular size, which we have estimated using data from \citet[figure 6: $z\sim1-3$ and $z > 3$ only]{Ikarashi2015}.  The source density parameter was calculated from the data in \citet{Cai2013} to be 0.011 arcsec$^{-2}$. The criterion for strong lens detection was simply that the 500\microns\ flux density be greater than the chosen flux limit, replacing the seeing and  signal-to-noise criteria of \citet{COLLETT2015}. A total of 10 per cent of the sky was simulated, to avoid Poisson errors in the models dominating the comparison with the data. A further simulation of $10\%$ of the sky was undertaken, from which a HERBS-like subset of source redshift $z_{\mathrm{source}}>2$ was selected. 

Figure \,\ref{fig:lenspop} shows the results of this modelling, compared to our observations. The flux density limit of $80$\,mJy applies to the targets in this paper, but we have included a $100$\,mJy prediction to demonstrate robustness to the precise choice of flux limit. The imposition of a HERBS-like $z_{\mathrm{source}}>2$ cut has a small effect on the magnification distribution. The HERBS-like selection (cyan in Fig.\,\ref{fig:lenspop}) agrees well with the observations out to magnifications of $\sim10$, but there are hints of an excess of objects at higher magnifications. Also shown are the magnifications of a 500\microns-selected sample measured in interferometric mapping by \citet{bussmann13}, and the magnifications estimated for South Pole Telescope targets that have 500\microns\ fluxes above $80$\,mJy and individual magnifications derived from lens modelling \citep{reuter20}, although note that this latter sample also has $1.4\,$mm and 870\microns\ flux cuts so is not straightforwardly comparable.

There are at least three potential explanations.  One possibility is that uncertainties in the magnification cause an Eddington-like bias in the magnification distribution in Fig.\,\ref{fig:lenspop}, and indeed \cite{aravena2016} find that these magnification estimates can have large uncertainties; interferometric mapping will help determine whether this is the case.  A second possible explanation is that this could indicate an additional contribution to the gravitational lensing optical depth that is not captured by the LensPop modelling. This model only considers galaxy-galaxy lensing, and has no contribution from strong lensing by galaxy clusters. There are other indications that the contribution from galaxy group or cluster lensing is not insignificant for wide blank-field surveys, as the measurement of a strong magnification bias \citep{gonzalez2017,dunne2020} or the fact that the submm--K-band offsets in \cite{bakx2020} are larger than the expectations from the modelling of \cite{Amvrosiadis2018}. Predicting the lensing optical depth from clusters from first principles is challenging, and there remain discrepancies with observations that may either be artefacts of simulation resolution or incorrect assumptions about the properties of dark matter \citep{Planelles2014,Meneghetti2020,Robertson2021}. If there are cluster or group lenses among the BEARS lensed sources, then the lensing geometries would need to be consistent with the angular size constraints from ALMA.  Indeed, while it is currently unknown which if any of our BEARS sample are cluster lenses, \cite{bakx2020} found that sources at lower flux densities are more likely to be lensed by cluster systems, in line with findings from \cite{gonzalez2017}. The possibility of proto-clusters will be discussed in future papers in this series.

A third possibility is that differential magnification affects the emission line {\it widths}, not simply the line luminosities (Section \ref{sec:diff}). The magnification estimates depend quadratically on line width (Fig.\,\ref{fig:L_v_V}), making magnification estimates sensitive to line width changes.  In support of this possibility, \cite{yang2017} found differences between the line width distributions of strongly and moderately lensed galaxies. The underlying line width distribution is purely a function of source properties, and therefore cannot depend on the foreground lensing galaxy.  Therefore, \citet{yang2017} attributed these differences to differential magnification affecting the line widths. 

However, unlike in \citet{yang2017}, we argue here that the line width differences need not on its own imply differential magnification in our sample. Even without differential magnification there are still selection effects at work in our sample that mean that the {\it observed} line width distribution could depend on magnification.  This is because it's relatively easy to find a rare highly-magnified example of a common faint galaxy, and it's relatively easy to find a rare bright galaxy with a common low magnification, but it's harder to find rare bright galaxies that also have rare high magnifications. Therefore, the {\it observed} magnification distribution might reasonably be expected to depend on the source luminosity, and therefore so would the offsets in Fig.\,\ref{fig:L_v_V}. The density of sources across this figure will also depend on the observational detection limits and the evolving luminosity function.

Therefore, we argue that a difference in our line width distributions of high-magnification and moderate-magnification objects is not enough {\it on its own} to prove that differential magnification of line widths is at work in our sample. The statistical effects on the population discussed above are incorporated in our LensPop simulations, though we consider only galaxy-galaxy lensing and do not consider the CO velocity and luminosity distributions within galaxies. The observations show a high-magnification excess above these models in Fig.\,\ref{fig:lenspop} that therefore may yet be attributable to differential magnification on line widths as argued by \citet{yang2017}, and indeed it would be hard to imagine differential magnification being unimportant in extreme high-magnification events. There are also observational precedents for line widths depending on magnification in spatially resolved systems \citep[e.g.,][]{yang2019,dye2015}. The large line widths in some systems suggest complex dynamics or mergers, which in turn suggests that line widths may be susceptible to differential magnification effects. 
It would be very useful to extend these simulations to consider the effects of differential magnification on line widths with the benefit of realistic dynamical models of the gas in SMGs, and to follow up the candidate high-magnification systems with higher resolution multi-wavelength imaging. 

\section{Conclusions}\label{sec:conclusions}

We present spectroscopic redshifts for a sample of the brightest ($S_{500} > 0.08$\,Jy) sources from the equatorial and Southern fields of the H-ATLAS Survey using data taken with the ALMA ACA and the full 12-m Array. From these observations, 71 robust redshift measurements were obtained from emission line detection using ALMA at a high efficiency (73 per cent of sources).  Our results can be further summarised as follows. 

\begin{enumerate}

\item Combining our results, we find that our sources lie within the redshift range $1.4 < z <4.5$ and, where available for comparison, all lie at redshifts greater than the likely foreground lens redshift.
\item  The distribution of CO emission line widths was found to be 100~km~s$^{-1}<$~FWHM~$<$~1290~km~s$^{-1}$, with a mean value of 494~km~s$^{-1}$, suggestive of complex dynamics.
\item The observed magnification distribution is consistent with galaxy-galaxy strong lensing models at magnifications $<10$, but there are hints of an excess at higher magnifications. This could be due to an additional contribution to the strong gravitational lensing optical depth in wide, blank-field surveys from galaxy groups and clusters as found in SMG magnification bias studies.

\end{enumerate}

\section*{Acknowledgements}
Herschel is an ESA space observatory with science instruments provided by European-led Principal Investigator consortia and with important participation from NASA.
SS was partly supported by the ESCAPE project; ESCAPE - The European Science Cluster of Astronomy \& Particle Physics ESFRI Research Infrastructures has received funding from the European Union's Horizon 2020 research and innovation programme under Grant Agreement no. 824064. SS also thanks the Science and Technology Facilities Council for financial support under grant ST/P000584/1. SU would like to thank the Open University School of Physical Sciences for supporting this work. We thank Z.-Y. Cai for kind assistance with the submm population models. RG thanks Churchill College in Cambridge for its hospitality and support.  TB and YT acknowledge funding from NAOJ ALMA Scientific Research Grant Numbers 2018-09B and JSPS KAKENHI No. 17H06130.  KK acknowledges the support by JSPS KAKENHI Grant Number 17H06130.  CY acknowledges support from an ESO Fellowship.  RI is funded by the Deutsche Forschungsgemeinschaft (DFG, German Research Foundation) under Germany's Excellence Strategy--EXC-2094--390783311.  JGN acknowledges the PGC 2018 project PGC2018-101948-B-100 (MICINN/FEDER).  MM acknowledges the support from grant PRIN MIUR 2017 - 20173ML3WW\_001.  This work benefited from the support of the project Z-GAL ANR-AAPG2019 of the French National Research Agency (ANR).  HD acknowledges financial support from the Spanish Ministry of Science, Innovation and Universities (MICIU) under the 2014 Ramón y Cajal program RYC-2014-15686, from the Agencia Estatal de Investigación del Ministerio de Ciencia e Innovación (AEI-MCINN) under grant (La evolución de los cíumulos de galaxias desde el amanecer hasta el mediodía cósmico) with reference (PID2019-105776GB-I00/DOI:10.13039/501100011033) and acknowledge support from the ACIISI, Consejería de Economía, Conocimiento y Empleo del Gobierno de Canarias and the European Regional Development Fund (ERDF) under grant with reference PROID2020010107.

This paper makes use of the following ALMA data: ADS/JAO.ALMA\#2016.2.00133.S, 2018.1.00804.S, and 2019.1.0147. ALMA is a partnership of ESO (representing its member states), NSF (USA) and NINS (Japan), together with NRC (Canada), MOST and ASIAA (Taiwan), and KASI (Republic of Korea), in cooperation with the Republic of Chile. The Joint ALMA Observatory is operated by ESO, AUI/NRAO and NAOJ. 

This research made use of Astropy,\footnote{http://www.astropy.org} a community-developed core Python package for Astronomy \citep{astropy2013,astropy2018}.  Data analysis made use of the Python package Numpy \citep{harris2020array}.
Some figures were made with the Python package Matplotlib \citep{Hunter:2007}.

\section*{Data Availability}

The {\it Herschel} SPIRE data can be downloaded at {\tt https://www.h-atlas.org}\,, while reduced and calibrated science-ready ALMA data is available on the ALMA Science Archive at {\tt https://almascience.eso.org/asax/}\,. This paper makes use of the following ALMA data:
ADS/JAO.ALMA\#2016.2.00133.S, 2018.1.00804.S, and
2019.1.01477.S.



\bibliographystyle{mnras}
\bibliography{main} 


\appendix
\section{Line properties, CO Luminosities and magnification factors}
In this Appendix we provide the identifications and line fluxes in table \ref{tab:lines}, while table \ref{tab:MagnificationFactors} provides the CO Luminosities, FWHM and magnification factors for all 71 galaxies.  

\onecolumn
\setlength\LTleft{-1cm}

\begin{longtable}{cccccccccc}
\caption{Identifications and line fluxes of the detected emission lines in Jy\,km\,s$^{-1}$. } 
\label{tab:lines} \\
\hline 
\multicolumn{1}{c}{H-ATLAS ID} 		 & \multicolumn{1}{c}{HerBS} 		 & \multicolumn{1}{c}{CO(2$-$1)} 		 & \multicolumn{1}{c}{CO(3$-$2)} 		  & \multicolumn{1}{c}{CO(4$-$3)} 		 & \multicolumn{1}{c}{CO(5$-$4)} 		 & \multicolumn{1}{c}{CO(6$-$5)}  		 & \multicolumn{1}{c} {CO(7$-$6)}		 & \multicolumn{1}{c}{CI(1$-$0)}		 & \multicolumn{1}{c} {H$_{2}$O $2_{11}-2_{02}$ }	 \\ 	
\hline \endhead			
\hline \multicolumn{10}{r}{\textit{Continued on next page}} \\
\endfoot
\hline
\endlastfoot

J012407.4$-$281434&11&-&11.0 $\pm$ 0.7& -& 21.1 $\pm$ 0.5& -&- & - &-\\
J013840.5$-$281856&14&-&-&7.3 $\pm$ 0.4&-&9.1 $\pm$ 0.5&-&2.1 $\pm$ 0.3&4.2 $\pm$ 0.5\\
J232419.8$-$323927&18&-&7.9 $\pm$ 0.7&9.0 $\pm$ 0.4&-&-&-&4.3 $\pm$ 0.5&-\\
J234418.1$-$303936&21&-&-&6.2 $\pm$ 0.7&-&10.8 $\pm$ 1.1&-&-&-\\
J002624.8$-$341738&22&-&-&-&13.3 $\pm$ 1.1&-&-&-&-\\
J004736.0$-$272951&24&-&5.3 $\pm$ 0.4&8.2 $\pm$ 0.5&-&-&-&5.0 $\pm$ 0.7&-\\
J235827.7$-$323244&25&-&6.6 $\pm$ 0.8&-&9.1 $\pm$ 0.4&-&-&-&-\\
J011424.0$-$33614&27&-&-&-&9.3 $\pm$ 0.6&-&10.7 $\pm$ 1.1&4.9 $\pm$ 2.1\footnotemark[1]&-\\ \footnotetext[1]{Observed transition is the CI(2-1)}
J230815.6$-$343801&28&-&-&5.3 $\pm$ 1.2&-&9.7 $\pm$ 0.7&-&-&-\\
J235623.1$-$354119&36&-&-&4.5 $\pm$ 0.5&6.7 $\pm$ 0.5&-&-&-&-\\
J232623.0$-$342642&37&-&3.5 $\pm$ 0.9&-&4.8 $\pm$ 0.5&-&-&-&-\\
J232900.6$-$321744&39&-&-&6.0 $\pm$ 0.6&-&-&-&-&-\\
J013240.0$-$330907&40&-&-&6.7 $\pm$ 0.5&-&-&-&-&-\\
J000124.9$-$354212&41&-&-&2.3 $\pm$ 0.4&-&-&4.8 $\pm$ 0.6&- &-\\
J000007.5$-$334060&42&-&-&3.2 $\pm$ 0.8&-&6.5 $\pm$ 0.4&-&-&-\\
J005132.8$-$301848&45&-&2.3 $\pm$ 0.4&-&-&-&-&3.0 $\pm$ 0.5&-\\
J225250.7$-$313658&47&-&4.6 $\pm$ 1.6&-&-&-&-&1.6 $\pm$ 0.3&-\\
J230546.3$-$331039&49A&-&3.5 $\pm$ 1.1&-&3.7 $\pm$ 0.2&-&-&-&-\\
&49B&-&-\footnotemark[1]\footnotetext[2]{The ACA flux is distributed between 49A \& 49B}&-&2.2 $\pm$ 0.4&-&-&-&-\\
J013951.9$-$321446&55&-&3.0 $\pm$ 0.4&-&6.7 $\pm$ 0.7&-&-&-&-\\
J003207.7$-$303724&56C&-&2.0 $\pm$ 0.3&-&2.5 $\pm$ 0.5&-&-&-&-\\
J004853.3$-$303110&57&-&-&6.4 $\pm$ 0.6&-&4.7 $\pm$ 0.5&-&-&-\\
J005724.2$-$273122&60&-&-&4.8 $\pm$ 0.5&-&-&-&-&-\\
J005132.0$-$302012&63&-&2.1 $\pm$ 0.3&-&-&-&-&2.0 $\pm$ 0.3&-\\
J223753.8$-$305828&68&-&6.5 $\pm$ 0.8&-&5.6 $\pm$ 0.5&-&-&-&-\\
J012416.0$-$310500&69A&-&5.5 $\pm$ 0.6&7.5 $\pm$ 0.4&-&-&-&3.0 $\pm$ 0.3&-\\
& 69B & -&0.6 $\pm$ 0.2&2.7 $\pm$ 0.4&-&-&-&0.7 $\pm$ 0.2&-\\
J012853.0$-$332719&73&-&-&-&5.8 $\pm$ 0.6&-&-&-&-\\
J005629.6$-$311206&77&-&5.5 $\pm$ 0.5&5.8 $\pm$ 0.6&-&-&-&-&-\\
J230002.6$-$315005&80A&-&3.6 $\pm$ 0.6&2.6 $\pm$ 0.5&-&-&-&-&-\\
&80B&-&-&1.7 $\pm$ 0.3&-&-&-&-&-\\
J002054.6$-$312752&81A&-&-&5.1 $\pm$ 0.8&-&-&-&-&-\\
&81B&-&1.8 $\pm$ 0.3&-&2.8 $\pm$ 0.4&-&-&-&-\\
J235324.7$-$331111&86&-&5.6 $\pm$ 0.5&-&5.2 $\pm$ 0.5&-&-&-&-\\
J005659.4$-$295039&90&-&-&3.3 $\pm$ 0.5&-&-&3.5 $\pm$ 0.7&4.1 $\pm$ 1.0&-\\
J234750.5$-$352931&93&-&4.1 $\pm$ 0.7&-&-&-&-&2.7 $\pm$ 0.5&-\\
J233024.1$-$325032&102&-&-&3.7 $\pm$ 0.6&-&4.9 $\pm$ 0.6&-&-&-\\
J225324.2$-$323504&103&-&-&-&7.1 $\pm$ 0.5&-&-&-&-\\
J001802.2$-$313505&106&-&4.2 $\pm$ 0.5&-&-&-&-&1.7 $\pm$ 0.3&-\\
J014520.0$-$313835&107&-&3.3 $\pm$ 0.4&-&3.1 $\pm$ 0.4&-&-&-&-\\
J223942.4$-$333304&111&-&6.6 $\pm$ 0.7&-&-&-&-&2.9 $\pm$ 0.3&-\\
J000806.8$-$351205&117&-&-&-&3.0 $\pm$ 0.4&-&5.4 $\pm$ 0.5&-&-\\
J012222.3$-$274456&120A&-&-&2.0 $\pm$ 0.6&4.5 $\pm$ 0.5&-&-&-&-\\
&120B&-&-&0.9 $\pm$ 0.2&3.8 $\pm$ 0.6&-&-&-&-\\
J223615.2$-$343301&121&-&-&4.6 $\pm$ 0.7&-&4.1 $\pm$ 0.4&-&-&-\\
J003717.0$-$323307&122A&-&-&-&3.3 $\pm$ 0.3&-&-&-&-\\
J233037.3$-$331218&123&-&5.9 $\pm$ 0.6&8.5 $\pm$ 0.5&-&-&-&4.6 $\pm$ 0.6&-\\
J225339.1$-$325550&131B&-&-&2.0 $\pm$ 0.3&-&-&-&4.5 $\pm$ 0.4&-\\
J231205.2$-$295027&132&-&4.0 $\pm$ 0.5&-&-&-&-&1.2 $\pm$ 0.2&-\\
J225611.7$-$325653&135A&-&-&-&2.7$\pm0.5$&-&-&1.1$\pm0.2$&-\\
J011730.3$-$320719&138B&1.2 $\pm$ 0.2&3.2 $\pm$ 0.5&-&-&-&-&-&-\\
J224759.7$-$310135&141&-&5.8 $\pm$ 1.0&7.4 $\pm$ 0.7&-&-&-&5.1 $\pm$ 0.8&-\\
J012335.1$-$314619&145&-&3.5 $\pm$ 0.5&-&5.3 $\pm$ 0.7&-&-&-&-\\
J232210.9$-$333749&146B&-&-&4.3 $\pm$ 0.5&-&-&-&-&-\\
J000330.7$-$321136&155&-&-&-&4.5 $\pm$ 0.3&-&-&-&-\\
J235122.0$-$332902&159A&-&4.7 $\pm$ 0.4&3.7 $\pm$ 0.4&-&-&-&-&-\\
&159B&-&0.8 $\pm$ 0.2&2.4 $\pm$ 0.4&-&-&-&-&-\\
J011014.5$-$314814&160&-&-&2.5 $\pm$ 0.4&-&2.9 $\pm$ 0.4&-&-&-\\
J000745.8$-$342014&163A&-&-&0.8 $\pm$ 0.2&1.3 $\pm$ 0.2&-&-&-&-\\
J225045.5$-$304719&168&-&3.3 $\pm$ 0.4&-&7.1 $\pm$ 0.9&-&-&-&-\\
J011850.1$-$283642&178A&-&3.0 $\pm$ 0.3&-&3.7 $\pm$ 0.5&-&-&-&-\\
&178B&-&2.6 $\pm$ 0.3&-&2.1 $\pm$ 0.5&-&-&-&-\\
&178C&-&1.4 $\pm$ 0.3&-&1.7 $\pm$ 0.3&-&-&-&-\\
J230538.5$-$312204&182&-&3.9 $\pm$ 0.7&4.0 $\pm$ 0.4&-&-&-&3.1 $\pm$ 0.5&-\\
J234955.7$-$330833&184&-&-&3.3 $\pm$ 0.3&-&-&-&2.1 $\pm$ 0.3&-\\
J225600.7$-$313232&189&-&-&4.7 $\pm$ 0.5&-&5.3 $\pm$ 0.6&-&-&-\\
J014313.2$-$332633&200&-&3.0 $\pm$ 0.5&3.7 $\pm$ 0.5&-&-&-&-&-\\
J005506.5$-$300027&207&7.0 $\pm$ 0.6&-&-&-&-&-&-&-\\
J225744.6$-$324231&208A&-&3.5 $\pm$ 0.7&-&-&-&-&2.0 $\pm$ 0.3&-\\
&208B&-&2.4 $\pm$ 0.4&-&-&-&-&-&-\\
J224920.6$-$332940&209&-&2.6 $\pm$ 0.4&3.3 $\pm$ 0.5&-&-&-&-&-\\
\hline
\end{longtable}

\onecolumn
\setlength\LTleft{+3cm}

\begin{longtable}{ccccc}

\caption{CO Luminosities, FWHM and magnification factors for all 71 galaxies. Magnification uncertainties are calculated using only the uncertainties in luminosity.} 
\label{tab:MagnificationFactors} \\
\hline 
\multicolumn{1}{c}{H-ATLAS ID} 		 & \multicolumn{1}{c}{HerBS ID} 	&\multicolumn{1}{c}{FWHM}	 & \multicolumn{1}{c}{$\mu L^{\prime}_{\rm CO(1-0)}$} 		 & \multicolumn{1}{c}{$\mu$} 		\\

& &	[\,km\,s$^{-1}$] & \multicolumn{1}{c}{[$10^{10}$\,K\,km\,s$^{-1}$\,pc$^{2}$]} 		 &  \\
\hline \endhead								
\hline \multicolumn{4}{r}{\textit{Continued on next page}} \\
\endfoot
\hline
\endlastfoot
J012407.4$-$281434&11&410 $\pm$ 80&78.1 $\pm$ 8.8&    18.4 $\pm$ 2.1\\ 
J013840.5$-$281856&14&270 $\pm$ 40&67.0 $\pm$ 8.2&    36.4 $\pm$ 4.5\\ 
J232419.8$-$323927&18&240 $\pm$ 40&40.7 $\pm$ 6.4&    27.9 $\pm$ 4.4\\ 
J234418.1$-$303936&21&550 $\pm$ 110&46.3 $\pm$ 6.8&   6.1$ \pm$ 0.9\\ 
J002624.8$-$341738&22&680 $\pm$ 150&70.3 $\pm$ 8.4&   6.1$ \pm$ 0.7\\ 
J004736.0$-$272951&24&490 $\pm$ 80&27.7 $\pm$ 5.3&    4.6$ \pm$ 0.9\\
J235827.7$-$323244&25&210 $\pm$ 40&55.6 $\pm$ 7.5&    49.8 $\pm$ 6.7\\ 
J011424.0$-$33614&27&500 $\pm$ 130&92.0 $\pm$ 6.5&    14.6 $\pm$ 1.0\\ 
J230815.6$-$343801&28&600 $\pm$ 100&51.9 $\pm$ 7.2&   5.7 $\pm$ 0.8 \\ 
J235623.1$-$354119&36&470 $\pm$ 100&30.0 $\pm$ 5.5&   5.4 $\pm$ 1.0  \\ 
J232623.0$-$342642&37&450 $\pm$ 90&25.0 $\pm$ 5.0&    4.9 $\pm$ 1.0 \\ 
J232900.6$-$321744&39&570 $\pm$ 160&42.0 $\pm$ 6.5&   5.1 $\pm$ 0.8 \\ 
J013240.0$-$330907&40&710 $\pm$ 80&20.6 $\pm$ 4.5&    1.6 $\pm$ 0.4 \\ 
J000124.9$-$354212&41A&680 $\pm$ 100&23.9 $\pm$ 4.9&  2.1 $\pm$ 0.4 \\ 
J000007.5$-$334060&42A&490 $\pm$ 90&23.6 $\pm$ 4.9&   3.9 $\pm$ 0.8 \\ 
J005132.8$-$301848&45A&210 $\pm$ 60&14.3 $\pm$ 3.8&   12.8 $\pm$ 3.4  \\ 
J225250.7$-$313658&47&180 $\pm$ 50&28.6 $\pm$ 5.3&    34.8 $\pm$ 6.4 \\ 
J230546.3$-$331039&49A&190 $\pm$ 20&26.3 $\pm$ 5.1&   28.8 $\pm$ 5.6 \\ 
&49B&470 $\pm$ 100&9.7 $\pm$ 3.2&                     1.7 $\pm$ 0.6  \\ 
J013951.9$-$321446&55&250 $\pm$ 60&21.7 $\pm$ 4.7&    13.7 $\pm$ 3.0 \\ 
J003207.7$-$303724&56C&580 $\pm$ 190&13.6 $\pm$ 3.7&  1.6 $\pm$ 0.4 \\ 
J004853.3$-$303110&57&350 $\pm$ 80&46.4 $\pm$ 6.8&    15.0 $\pm$ 2.2 \\ 
J005724.2$-$273122&60&380 $\pm$ 90&34.7 $\pm$ 5.9&    9.5 $\pm$ 1.6 \\ 
J005132.0$-$302012&63A&400 $\pm$ 60&12.4 $\pm$ 3.5&   3.1 $\pm$ 0.9 \\ 
J223753.8$-$305828&68&430 $\pm$ 60&48.9 $\pm$ 7.0&    10.5 $\pm$ 1.5 \\ 
J012416.0$-$310500&69A&500 $\pm$ 130&25.9 $\pm$ 5.1&  4.1 $\pm$ 0.8 \\ 
&69B&120 $\pm$ 10&2.83 $\pm$ 1.7&                     7.7 $\pm$ 4.6 \\ 
J012853.0$-$332719&73&630 $\pm$ 140&31.3 $\pm$ 5.6&   3.1 $\pm$ 0.6 \\ 
J005629.6$-$311206&77&910 $\pm$ 150&28.9 $\pm$ 5.4&   1.4 $\pm$ 0.3 \\ 
J230002.6$-$315005&80A&650 $\pm$ 30&19.3 $\pm$ 4.4&   1.8 $\pm$ 0.4 \\ 
&80B&340 $\pm$ 70&5.2 $\pm$ 2.3&                      1.8 $\pm$ 0.8  \\ 
J002054.6$-$312752&81A&670 $\pm$ 200&35.0 $\pm$ 5.9&  3.1 $\pm$ 0.5 \\ 
&81B&650 $\pm$ 190&12.4 $\pm$ 3.5&                    1.2 $\pm$ 0.3 \\ 
J235324.7$-$331111&86&560 $\pm$ 120&38.0 $\pm$ 6.2&   4.8 $\pm$ 0.8 \\ 
J005659.4$-$295039&90A&580 $\pm$ 150&33.0 $\pm$ 5.7&  3.9 $\pm$ 0.7 \\ 
J234750.5$-$352931&93&640 $\pm$ 160&24.9 $\pm$ 5.0&   2.4 $\pm$ 0.5 \\ 
J233024.1$-$325032&102&360 $\pm$ 120&27.1 $\pm$ 5.2&  8.3 $\pm$ 1.6 \\ 
J225324.2$-$323504&103&580 $\pm$ 130&36.7 $\pm$ 6.1&  4.3 $\pm$ 0.7 \\ 
J001802.2$-$313505&106A&500 $\pm$ 170&25.0 $\pm$ 5.0& 4.0 $\pm$ 0.8 \\ 
J014520.0$-$313835&107&230 $\pm$ 50&22.2 $\pm$ 4.7&  16.6 $\pm$ 3.5 \\ 
J223942.4$-$333304&111&600 $\pm$ 120&39.3 $\pm$ 6.3&  4.3 $\pm$ 0.7 \\ 
J000806.8$-$351205&117A&660 $\pm$ 170&29.8 $\pm$ 5.5& 2.7 $\pm$ 0.5 \\ 
J012222.3$-$274456&120A&480 $\pm$ 150&13.5 $\pm$ 3.7& 2.3 $\pm$ 0.6 \\ 
&120B&740 $\pm$ 190&21.0 $\pm$ 4.6&                   1.5 $\pm$ 0.3 \\ 
J223615.2$-$343301&121A&360 $\pm$ 80&41.4 $\pm$ 6.4& 12.7 $\pm$ 2.0 \\ 
J003717.0$-$323307&122A&360 $\pm$ 90&16.5 $\pm$ 4.1&  5.1 $\pm$ 1.3 \\ 
J233037.3$-$331218&123&280 $\pm$ 80&30.1 $\pm$ 5.5&  15.2 $\pm$ 2.8\\ 
J225339.1$-$325550&131B&750 $\pm$ 130&7.1 $\pm$ 2.7&  0.5 $\pm$ 0.2 \\ 
J231205.2$-$295027&132&460 $\pm$ 100&25.6 $\pm$ 5.1&  4.8 $\pm$ 1.0 \\ 
J225611.7$-$325653&135A&380 $\pm$ 60&16.5 $\pm$ 4.1&  4.5 $\pm$ 1.1 \\ 
J011730.3$-$320719&138B&110 $\pm$ 20&3.9 $\pm$ 2.0&  12.7 $\pm$ 6.5  \\ 
J224759.7$-$310135&141&370 $\pm$ 90&25.3 $\pm$ 5.0&   7.3 $\pm$ 1.4 \\ 
J012335.1$-$314619&145A&730 $\pm$ 110&25.8 $\pm$ 5.1& 1.9 $\pm$ 0.4 \\ 
J232210.9$-$333749&146B&400 $\pm$ 80&13.7 $\pm$ 3.7&  3.4 $\pm$ 0.9 \\ 
J000330.7$-$321136&155A&500 $\pm$ 150&25.1 $\pm$ 5.0& 4.0 $\pm$ 0.8 \\ 
J235122.0$-$332902&159A&280 $\pm$ 80&24.9 $\pm$ 5.0& 12.6 $\pm$ 2.5   \\ 
&159B&330 $\pm$ 90&4.30 $\pm$ 2.1&                     1.6 $\pm$ 0.8  \\ 
J011014.5$-$314814&160&350 $\pm$ 70&25.1 $\pm$ 5.0&    8.1 $\pm$ 1.6 \\ 
J000745.8$-$342014&163A&470 $\pm$ 150&5.44 $\pm$ 2.3&  1.0 $\pm$ 0.4 \\ 
J225045.5$-$304719&168&720 $\pm$ 130&22.9 $\pm$ 4.8&   1.8 $\pm$ 0.4 \\ 
J011850.1$-$283642&178A&680 $\pm$ 190&21.7 $\pm$ 4.7&  1.9 $\pm$ 0.4  \\ 
&178B&710 $\pm$ 120&18.8 $\pm$ 4.3&                    1.5 $\pm$ 0.3 \\ 
&178C&610 $\pm$ 250&10.1 $\pm$ 3.2&                    1.1 $\pm$ 0.3 \\ 
J230538.5$-$312204&182&870 $\pm$ 170&20.4 $\pm$ 4.5&   1.1 $\pm$ 0.2 \\ 
J234955.7$-$330833&184&320 $\pm$ 70&15.5 $\pm$ 4.0&    6.0 $\pm$ 1.5 \\ 
J225600.7$-$313232&189&310 $\pm$ 90&34.8 $\pm$ 5.9&   14.3 $\pm$ 2.4 \\ 
J014313.2$-$332633&200&1290 $\pm$ 400&15.1 $\pm$ 3.9&  0.4 $\pm$ 0.1  \\ 
J005506.5$-$300027&207&460 $\pm$ 70&27.8 $\pm$ 5.3&    5.2 $\pm$ 1.0 \\ 
J225744.6$-$324231&208A&770 $\pm$ 140&22.5 $\pm$ 4.7&  1.5 $\pm$ 0.3 \\ 
                  &208B&720 $\pm$ 160&15.5 $\pm$ 3.9&  1.2 $\pm$ 0.3 \\ 
J224920.6$-$332940&209A&500 $\pm$ 30&13.8 $\pm$ 3.7&   2.2 $\pm$ 0.6 \\

\end{longtable}

\bsp	
\label{lastpage}
\end{document}